\documentclass[
 reprint,
 superscriptaddress,
 amsmath,amssymb,
 aps,
 floatfix,
]{revtex4-1}

\usepackage[pdftex]{graphicx}% Include figure files
\usepackage{epstopdf}
\usepackage{subfigure}
\usepackage{dcolumn}% Align table columns on decimal point
\usepackage{bm}% bold math
\usepackage{verbatim}
\begin{document}

\preprint{APS/123-QED}

\title{Complete modes of star-shaped oscillating drops}

\author{Jiahao Dong}
%\email{161120020@smail.nju.edu.cn}
\affiliation{
School of Physics, Nanjing University, Nanjing 210093, China
}

\author{Yuping Liu}
%\email{lypever@gmail.com}
\affiliation{
Kuang Yaming Honors School, Nanjing University, Nanjing 210093, China
}

\author{Qian Xu}
\email{qian.xu@smail.nju.edu.cn}
\affiliation{
School of Physics, Nanjing University, Nanjing 210093, China
}

\author{Yinlong Wang}
%\email{wangylphy@nju.edu.cn}
\affiliation{
School of Physics, Nanjing University, Nanjing 210093, China
}

\author{Sihui Wang}
\email{wangsihui@nju.edu.cn}
\affiliation{
School of Physics, Nanjing University, Nanjing 210093, China
}

\date{\today}

\begin{abstract}

The star-shaped oscillation of water drops has been observed in various physical situations with different sources of vertical excitation. Previous studies apply a quasi-2D model to analyze the resonance frequency and only consider the azimuthal oscillation mode. In this paper, we find that the upper surface of water drops also develop different motion patterns due to parametric instability and it is the coupling of surface motion and azimuthal oscillation that leads to star-shaped oscillations. We will introduce the analysis of the surface mode, combining with the azimuthal mode to give a complete description of the motion of water drops. We propose a new dispersion relation based on the complete description, which provides a significant increase of accuracy in predicting the oscillating frequencies.

%\begin{description}
%\end{description}

\end{abstract}

\pacs{Valid PACS appear here}% PACS, the Physics and Astronomy
\maketitle

%\tableofcontents

\section{Introduction}\label{Introduction}

Liquid drops driven by vertical excitation can develop star-shaped oscillations under certain conditions, including drops on a vertically vibrating hydrophobic plate~\cite{PRL-Triplon-Modes-of-Puddles}, drops on a pulsating air cushion~\cite{PRL-Gas-Film-Levitated-Liquids-Shape-Fluctuations-of-Viscous-Drops}, acoustically levitating drops~\cite{PRE-Parametrically-excited-sectorial-oscillation-of-liquid-drops-floating-in-ultrasound}, metal drops placed on an oscillating magnetic field~\cite{JFM-Free-surface-horizontal-waves-generated-by-low-frequency-alternating-magnetic-fields}. These oscillation systems are considered to undergo parametric resonance because the water drops are observed to oscillate in half of the frequency of the excitation source experimentally~\cite{JPSJ-Self-Induced-Vibration-of-a-Water-Drop-Placed-on-an-Oscillating-Plate,EPJST-Star-drops-formed-by-periodic-excitation-and-on-an-air-cushion-A-short-review,PRE-Parametrically-excited-sectorial-oscillation-of-liquid-drops-floating-in-ultrasound}. Even in the Leidenfrost effect~\cite{Annual-Review-of-Fluid-Mechanics-Leidenfrost-Dynamics,JPSJ-Vibration-of-a-Flattened-Drop.-I.-Observation,JPSJ-Vibration-of-a-Flattened-Drop.-II.-Normal-Mode-Analysis,JASA-Vibrations-of-Evaporating-Liquid-Drops,Physica-A-Nitrogen-stars-morphogenesis-of-a-liquid-drop} that the water drops are levitated by water vapor on a very hot plate, recent studies have found the periodically vibrating pressure of the vapor layer under the water drops can serve as parametric excitation~\cite{PRF-Star-shaped-oscillations-of-Leidenfrost-drops,Physics-of-Fluids-The-many-faces-of-a-Leidenfrost-drop}. Despite these experimental evidences, the mechanism that induces the parametric instability is still unclear.

In the past, the oscillation modes of the flattened water drops are described by Rayleigh equation, proposed for inviscid cylindrical drops. The eigen frequencies for small deformations are given by~\cite{Rayleigh-On-the-Capillary-Phenomena-of-Jets,JPSJ-Vibration-of-a-Flattened-Drop.-II.-Normal-Mode-Analysis}:
\begin{align}
\omega_n&=\sqrt{\frac{n(n^2-1)\sigma}{\rho R^3}}\label{Rayleigh-equation}
\end{align}
where $n$ and $\omega_n$ represent azimuthal mode and corresponding eigen frequency respectively, and $R$ is the balanced radius of the water drop, $\sigma$ the surface tension constant, $\rho$ the density of water. In the equation, the motion of the water drops is simplified as two-dimensional. Though the dispersion relation~\ref{Rayleigh-equation} gives right trend between the oscillation mode number $n$ and the frequency, more detailed studies by Bouwhuis et al.~\cite{PRE-Oscillating-and-star-shaped-drops-levitated-by-an-airflow} found that the measured resonance frequencies of star-shaped drops levitating on a steady ascending airflow are lower than that given by~\ref{Rayleigh-equation}, overestimating of the stiffness coefficient of the oscillating drop.

The two-dimensional model apparently oversimplifies the oscillation modes and loses accuracy when upper surface motion is appreciable. Shen et al. have already found that the upper surface develops petal-shaped motion patterns~\cite{PRE-Parametrically-excited-sectorial-oscillation-of-liquid-drops-floating-in-ultrasound,JFM-Free-surface-horizontal-waves-generated-by-low-frequency-alternating-magnetic-fields,Observation-of-the-shape-of-a-water-drop-on-an-oscillating-Teflon-plate}. In our experiments, we also find that, with the increase of mode number $n$, the upper surface of the water drop becomes unstable and forms various patterns.

In this paper, we will present a complete theoretical model to derive the surface normal modes of the drop and combine the azimuthal oscillation modes to give a complete description of the motion of water drops. We verify that the vertical driving force induces an instability on the upper surface due to parametric resonance. The coupling of the surface motion and azimuthal oscillation leads to the appearance of star-shaped drops oscillating at half of the driving frequency. The surface oscillation will be described in much the way as the Faraday waves. Faraday waves are patterns of standing waves observed at the free surface, when a layer of liquid in a container is driven to vibrate vertically~\cite{Faraday-Waves,Rayleigh-On-maintained-vibrations,The-Royal-Society-The-stability-of-the-plane-free-surface-of-a-liquid-in-vertical-periodic-motion,RSI-Patterning-of-particulate-films-using-Faraday-waves}. 

We will introduce a new mode number $m$, the surface mode number, and propose a new dispersion relation, in which the oscillation frequency of the water drop depends on both mode numbers $n$ and $m$. The stiffness coefficient is reduced due to the additional surface mode, so that the eigen frequencies are lower than that~\ref{Rayleigh-equation} predicts. We also conduct experiments by exciting star-shaped oscillating drops on a hydrophobic vibrating substrate. Our dispersion relation can predict the oscillation frequencies with high accuracy.

\section{Theory}\label{Model}

%%% figure 1

\begin{figure*}[t]
\centering

\subfigure[n=5]{
\begin{minipage}[t]{0.23\textwidth}
\centering
\includegraphics[width=1\textwidth,height=0.8\textwidth]{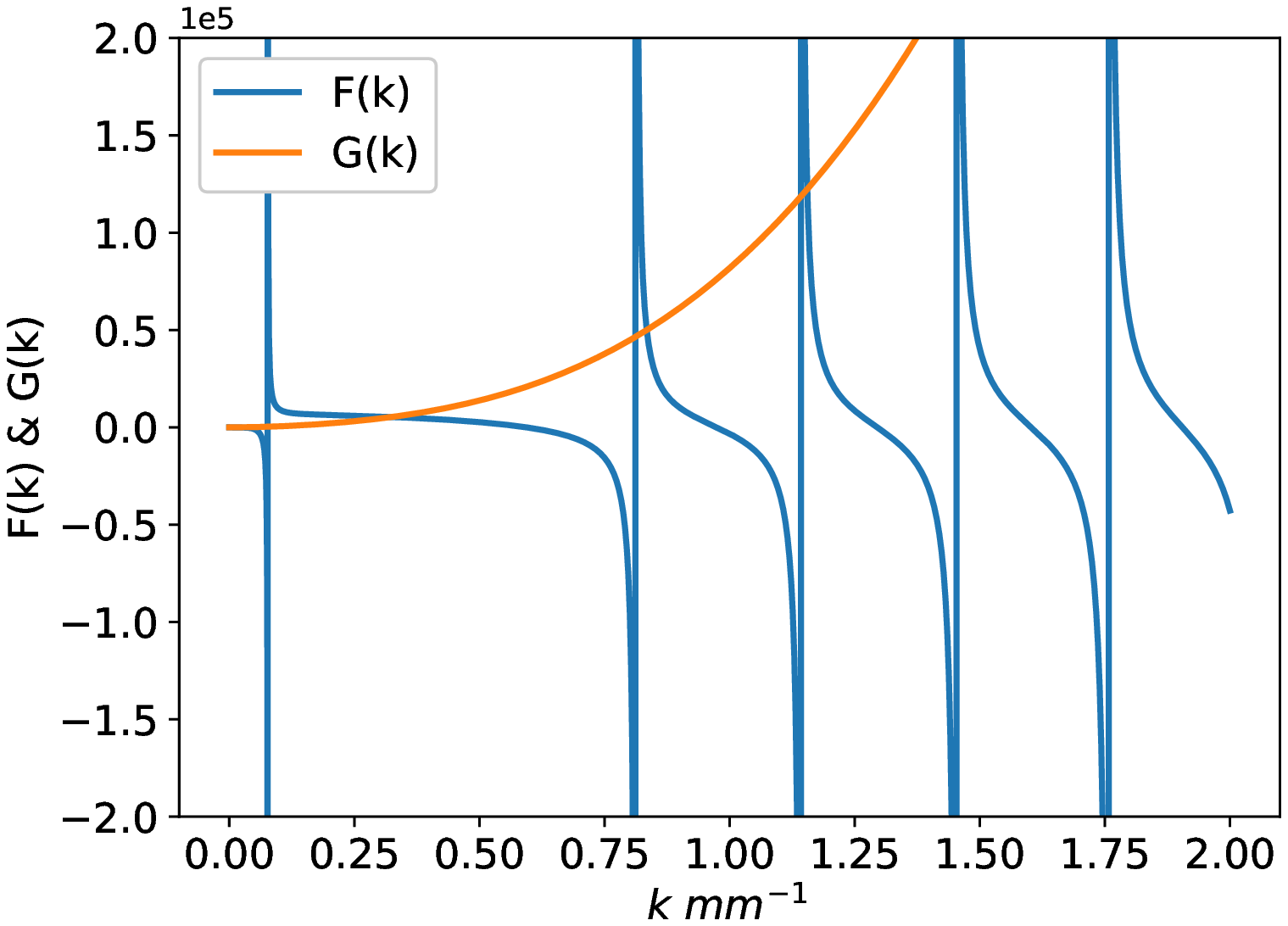}
\end{minipage}
}
\subfigure[n=6]{
\begin{minipage}[t]{0.23\textwidth}
\centering
\includegraphics[width=1\textwidth,height=0.8\textwidth]{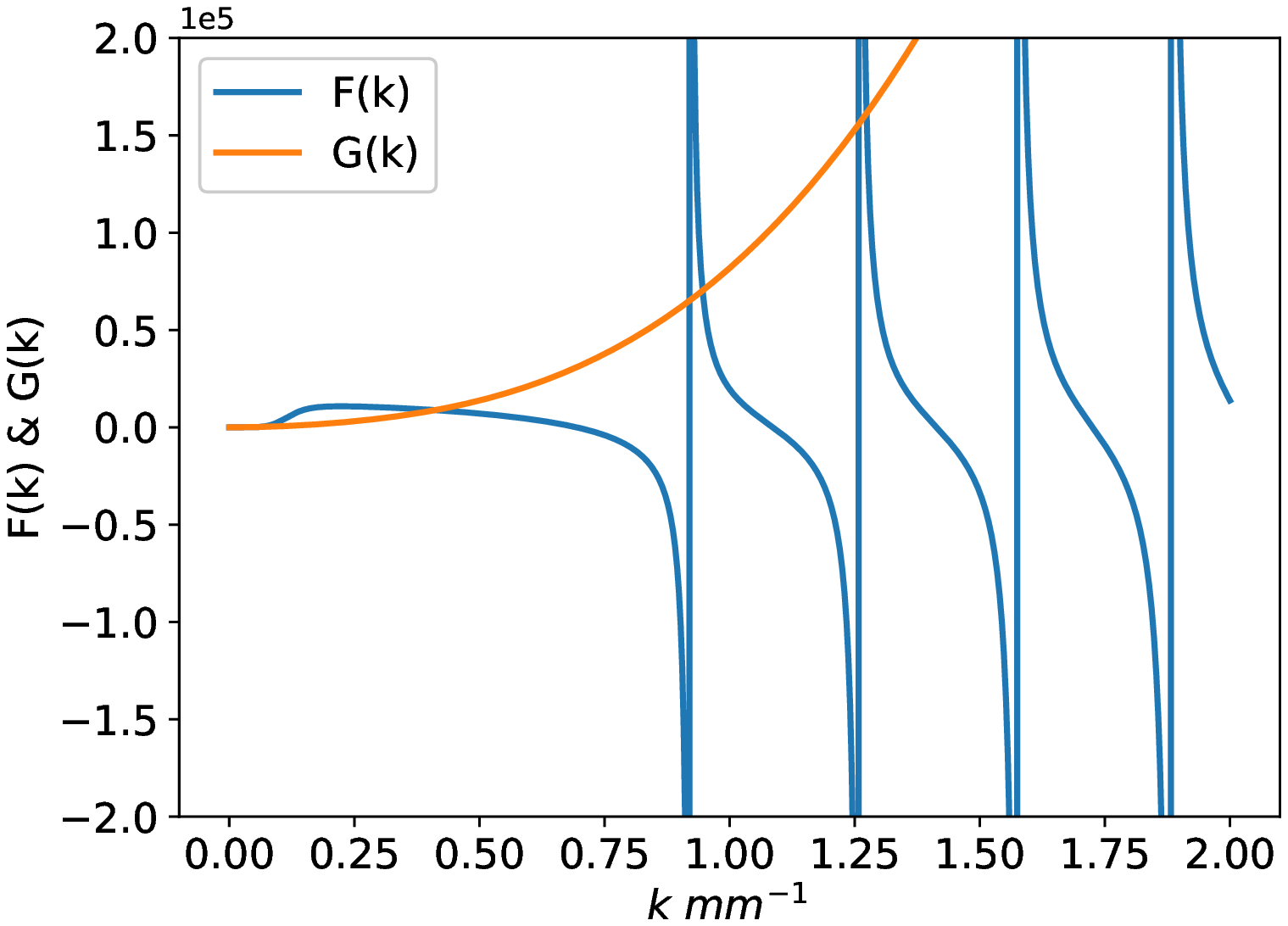}
\end{minipage}
}
\subfigure[n=7]{
\begin{minipage}[t]{0.23\textwidth}
\centering
\includegraphics[width=1\textwidth,height=0.8\textwidth]{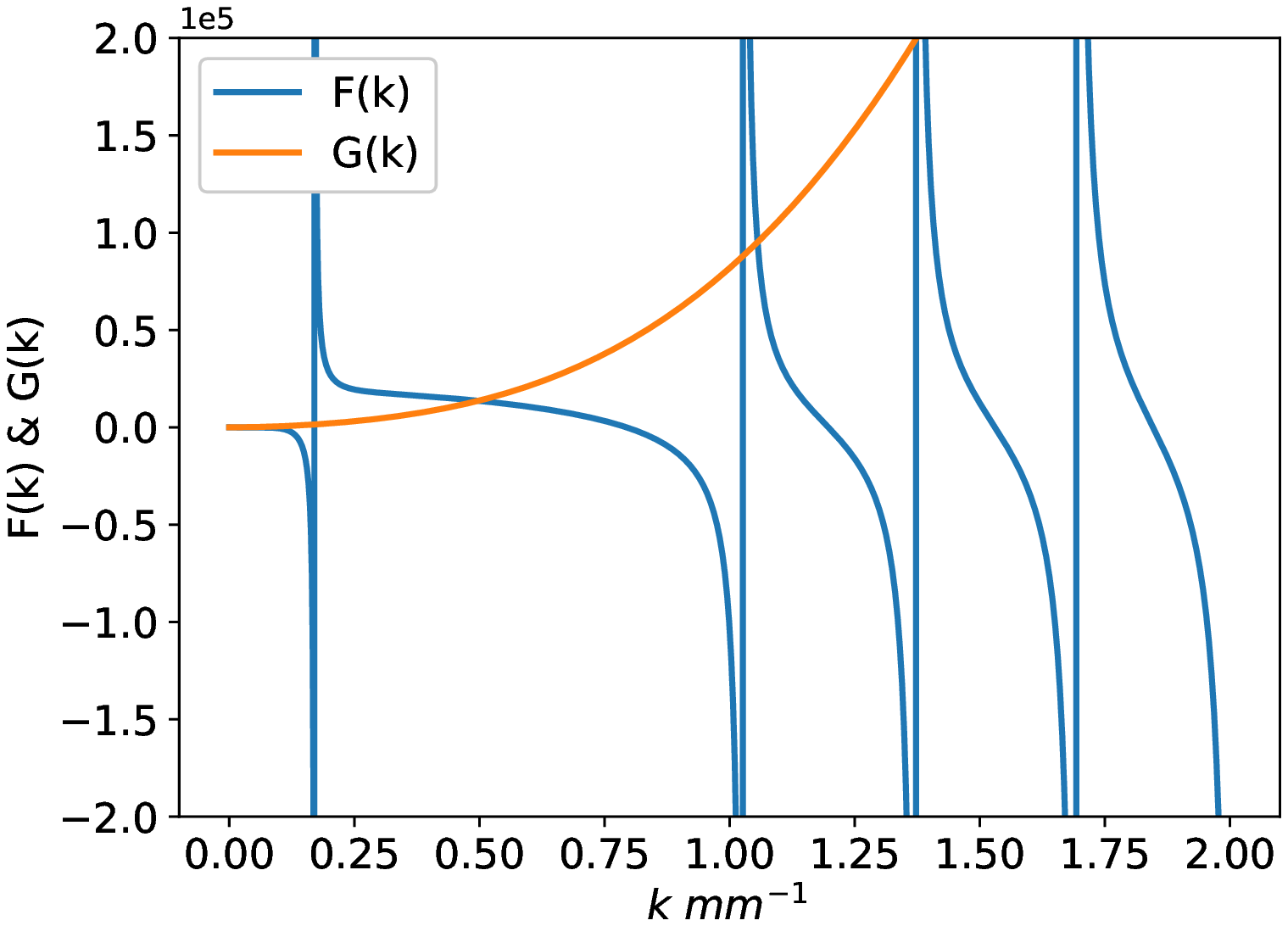}
\end{minipage}
}
\subfigure[n=8]{
\begin{minipage}[t]{0.23\textwidth}
\centering
\includegraphics[width=1\textwidth,height=0.8\textwidth]{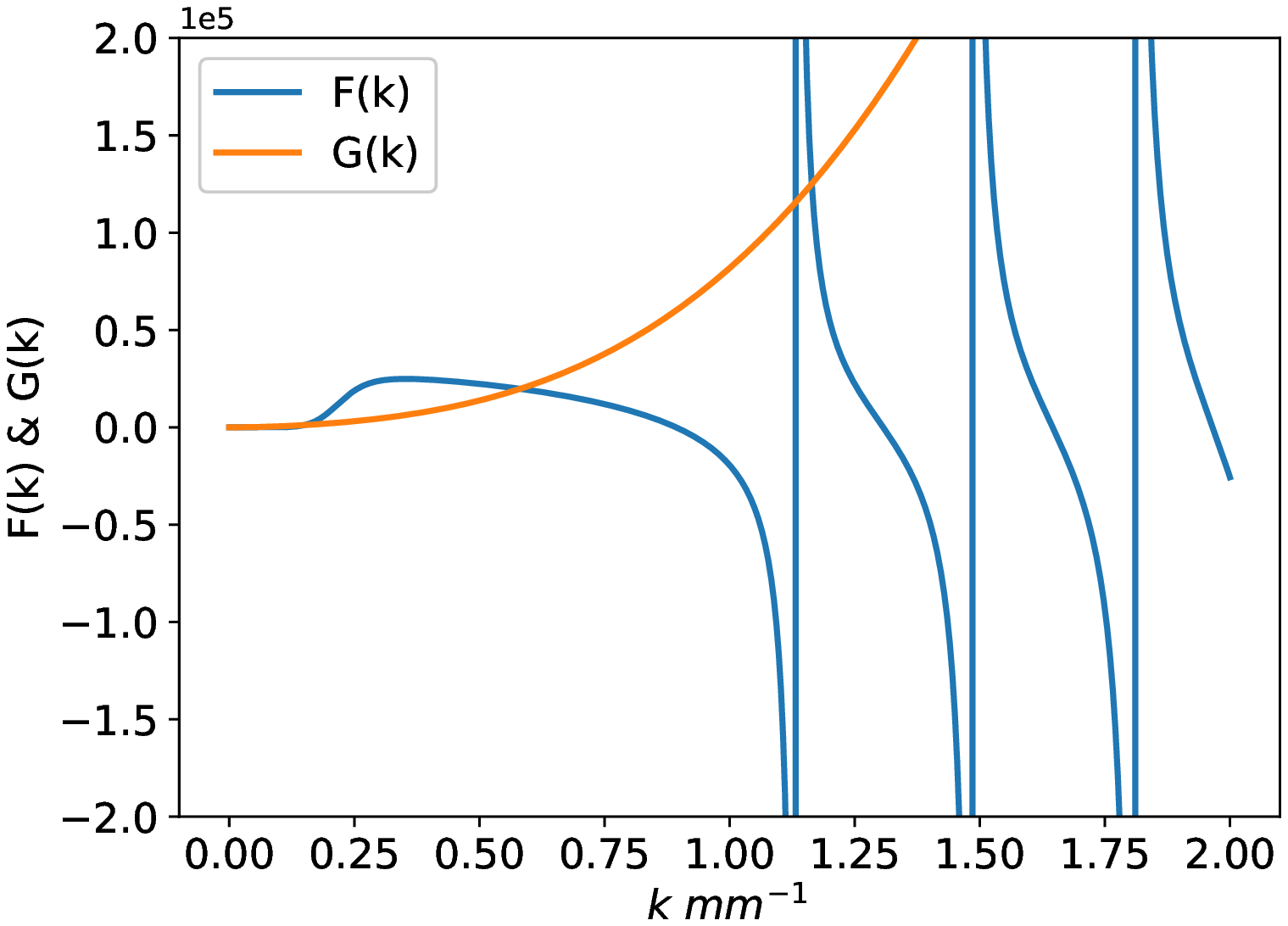}
\end{minipage}
}

\caption{$F(k)$ and $G(k)$ are plotted with orange and blue lines respectively, whose intersections determine the eigen values of $k$. From (a) to (d) azimuthal mode $n$ increases from 5 to 8. $R=10.8mm$($2l_c$), $\sigma=72mN/m$($25^{\circ} C$).}
\label{Theoretical eigen value calculation}
\end{figure*}

%%% figure 1

We consider the oscillation of water drops on a hydrophobic substrate, which vibrates periodically and serves as a vertical excitation. A water drop on a stable hydrophobic substrate is approximately a flattened cylinder with thickness about twice the capillary length ($2l_c$), determined by a balance between the surface tension and the gravity~\cite{Physics-of-Fluids-Leidenfrost-drops}.

We take cylindrical coordinates $(r,\theta,z)$ with respect to the frame of reference moving with the vibrating substrate, in which the undisturbed upper surface and periphery of the drop are $z=H$ and $r=R$ respectively. For an oscillating star-shaped drop, we use $h(r,\theta,t)$ and $a(z,\theta,t)$ to describe the small deviation of the upper surface and the periphery, hence the boundaries of the drop are described using $z=H+h$, $r=R+a$.

For irrotational liquid the velocity potential $\phi(r,\theta,z,t)$ ~\cite{An-Introduction-to-Fluid-Dynamics} satisfies Eulerian equation
\begin{align}
\frac{\partial\phi}{\partial t}+\frac{1}{2}(\nabla\phi)^2+\frac{p}{\rho}+
U=0\label{Eulerian-equation}
\end{align}
where $\nabla\phi$ is the fluid velocity, $p$ is the internal pressure, $U$ is the potential of external forces. For small movements of the water drop, we linearize the equation by omitting the term $(\nabla\phi)^2$.

The pressure at the free liquid surface~\cite{Hydrodynamics} is
\begin{align}
p= \sigma(\gamma_1+\gamma_2)\label{Young-Laplace}
\end{align}
where $\sigma$ is the surface tension and $\gamma_1$, $\gamma_2$ are the principal radius of curvature (ROC) of the surface. 

The ROCs of the upper surface can be obtained from membrane theory~\cite{rayleigh1894theory}, by which we obtain $p=\sigma\nabla^2h$. So the boundary condition of the upper surface is
\begin{align}
\left(\frac{\partial\phi}{\partial t}\right)_{z=H+h}+\frac{\sigma}{\rho}\nabla^2h+\left(g-Fcos(\Omega t)\right)h=0\label{surface-boundary-condition}
\end{align}

Applying~\ref{Young-Laplace} to the periphery, we can get the boundary condition as
\begin{align}
\left(\frac{\partial\phi}{\partial t}\right)_{r=R+a}-\frac{\sigma}{\rho R^2}\left(a+\frac{\partial^2a}{\partial\theta^2}\right)=0\label{periphery-boundary-condition}
\end{align}

The bottom surface is in contact with the hydrophobic surface, thus the boundary condition is
\begin{equation}
\left(\frac{\partial \phi}{\partial z}\right)_{z = 0} = 0
\label{bottom-equation-of-motion}
\end{equation}

Furthermore, at the liquid surfaces, $z=H+h$ and $r=R+a$, there are kinematic boundary conditions
\begin{align}
\left(\frac{\partial \phi}{\partial z}\right)_{z=H+h}&=\frac{\partial h}{\partial t}\label{kinematic1}\\
\left(\frac{\partial \phi}{\partial r}\right)_{r=R+a}&=\frac{\partial a}{\partial t}\label{kinematic2}
\end{align}

Considering the incompressibility, the equation of continuity can be written as
\begin{align}
\frac{\partial^2\phi}{\partial r^2}+\frac{1}{r}\frac{\partial\phi}{\partial r}+
\frac{1}{r^2}\frac{\partial^2\phi}{\partial r^2}+\frac{\partial^2\phi}{\partial z^2}=0\label{laplace-equation}
\end{align}

For small deformations, Eq.~\ref{laplace-equation} and Eqs.~\ref{surface-boundary-condition},~\ref{periphery-boundary-condition},~\ref{bottom-equation-of-motion} are variable-separable. The oscillation of a flattened cylinder drop can be decomposed on the Bessel basis. Hence the general solution of the velocity potential is
\begin{align}
\phi&=\sum_{k,n}cosh(kz)J_n(kr)sin(n\theta)A_{k,n}(t)\label{phi}
\end{align}
where $J_n$ refers to Bessel function. Here $A_{k,n}(t)$ represents the time-related coefficient. Similarly, $h(r,\theta,t)$ and $a(z,\theta,t)$ can be expressed as
\begin{align}
h=\sum_{k,n}J_n(kr)sin(n\theta)B_{k,n}(t)\label{h}\\
a=\sum_{k,n}cosh(kz)sin(n\theta)C_{k,n}(t)\label{a}
\end{align}
Inserting Eqs.~\ref{phi},~\ref{h} and~\ref{a} into Eqs.~\ref{kinematic1} and~\ref{kinematic2}, the coefficients $A_{k,n}(t)$, $B_{k,n}(t)$ and $C_{k,n}(t)$ satisfy
\begin{align}
\frac{dC_{k,n}(t)}{dt}&=kJ'_n (kR)A_{k,n}(t)\label{time-relation-1}\\
\frac{dB_{k,n}(t)}{dt}&=ksinh(kH)A_{k,n}(t)\label{time-relation-2}
\end{align}

%%% figure 2

\begin{figure}[t]
\centering

\subfigure[m=1]{
\begin{minipage}[t]{0.22\textwidth}
\centering
\includegraphics[width=1\textwidth,height=0.8\textwidth]{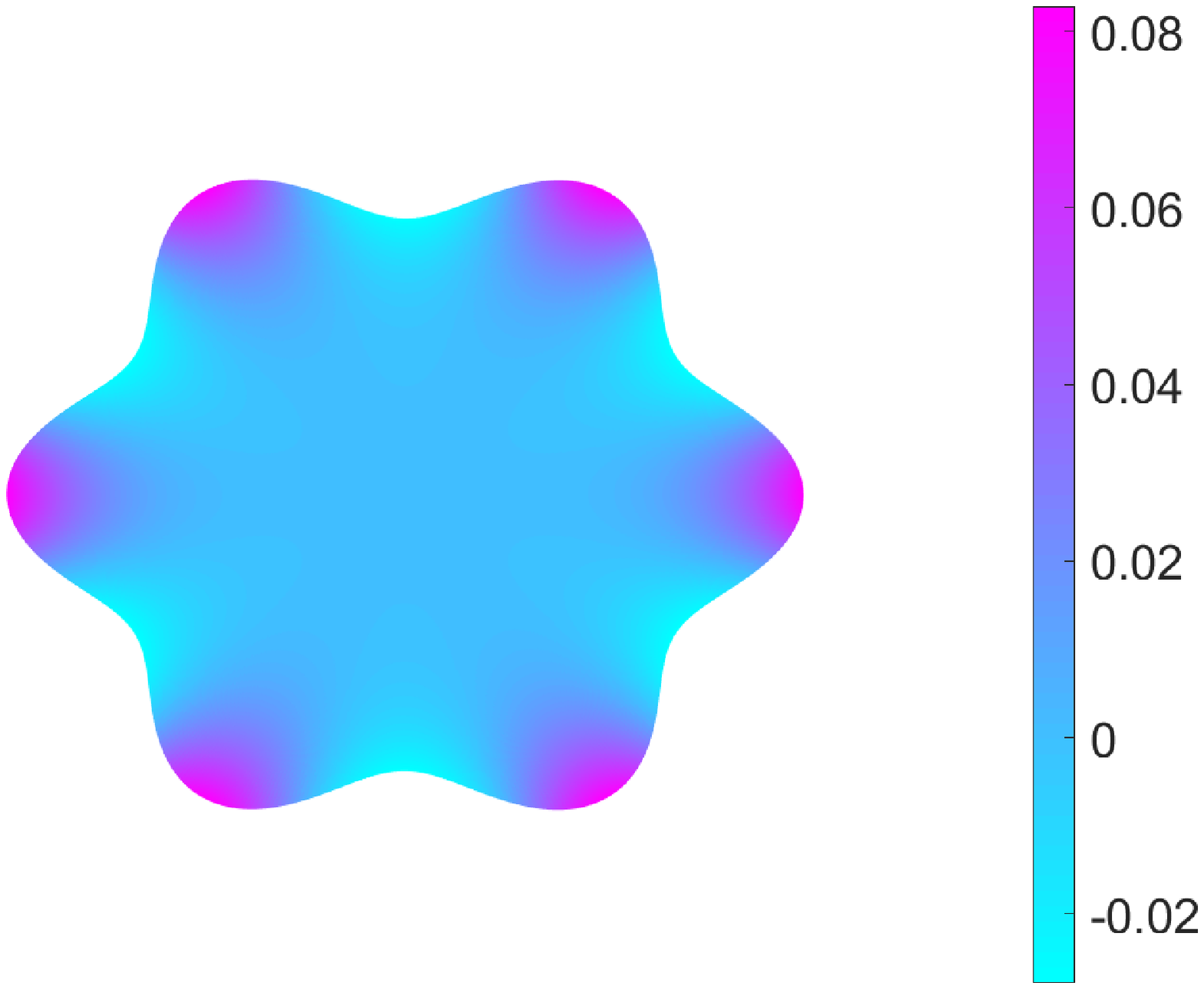}
\end{minipage}
}
\subfigure[m=2]{
\begin{minipage}[t]{0.22\textwidth}
\centering
\includegraphics[width=1\textwidth,height=0.8\textwidth]{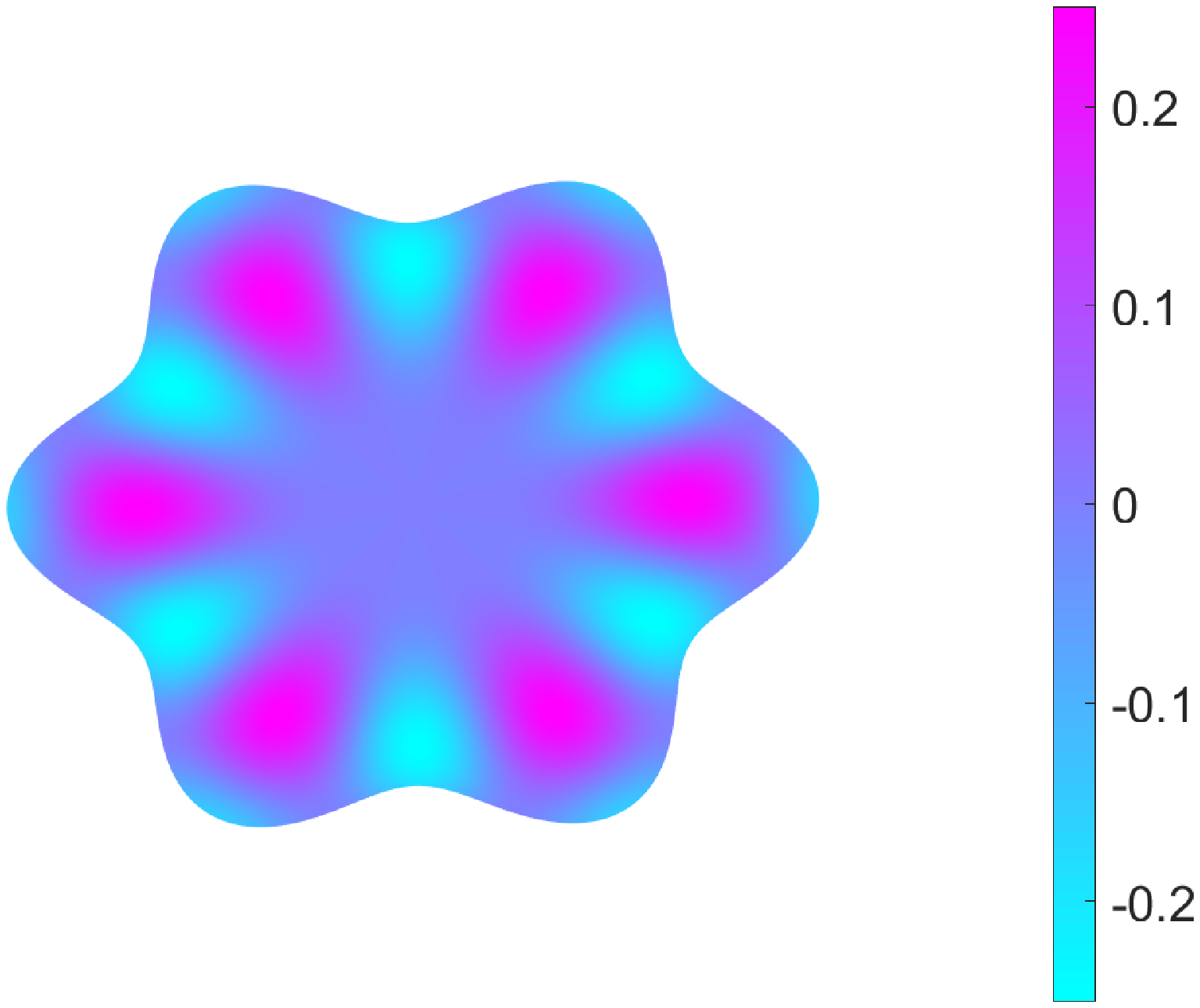}
\end{minipage}
}

\subfigure[m=3]{
\begin{minipage}[t]{0.22\textwidth}
\centering
\includegraphics[width=1\textwidth,height=0.8\textwidth]{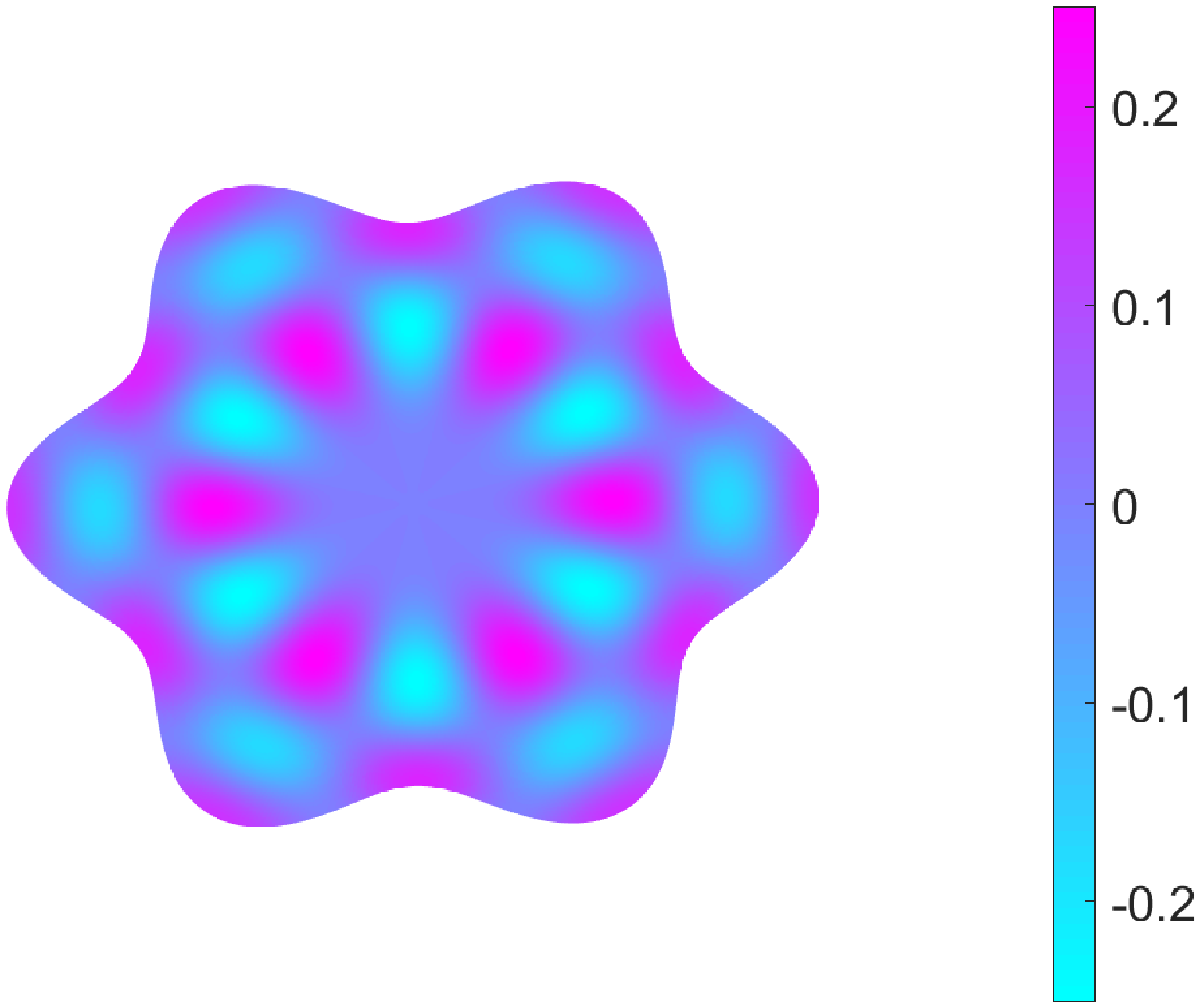}
\end{minipage}
}
\subfigure[m=4]{
\begin{minipage}[t]{0.22\textwidth}
\centering
\includegraphics[width=1\textwidth,height=0.8\textwidth]{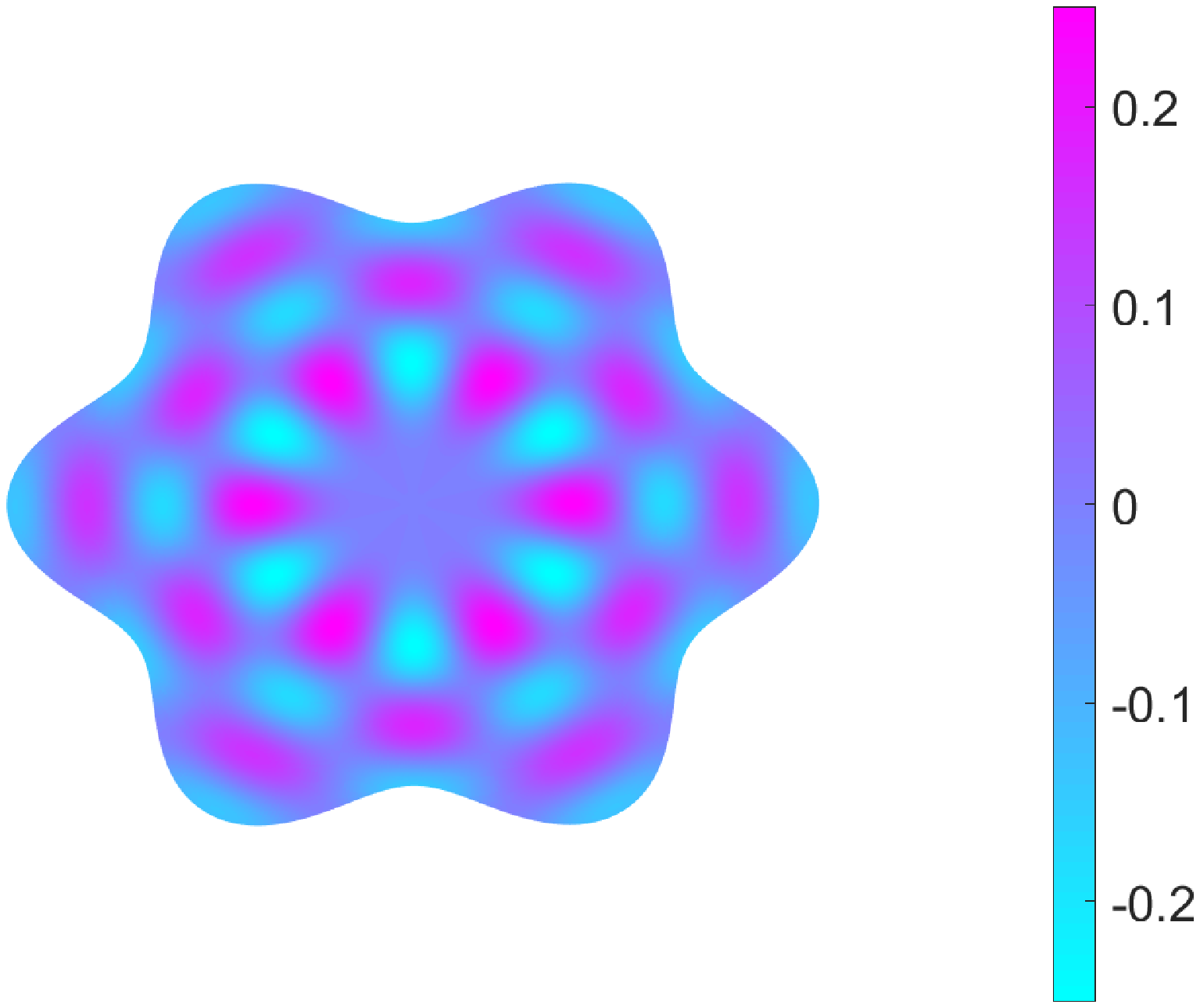}
\end{minipage}
}

\caption{Top views of the surface motion patterns (first four modes). The colormap of the surface corresponds to the height variation. The colorbars represent height (in $mm$). The corresponding values of $k_m$ are $k_1=0.402mm^{-1}$, $k_2=0.915mm^{-1}$, $k_3=1.258mm^{-1}$ and $k_4=1.574mm^{-1}$. $n=6$, $R=10.8mm$ ($2l_c$) and $\sigma=72mN/m$ ($25^{\circ} C$).}
\label{Surfaces}
\end{figure}
%%% figure 2

Then substituting Eqs.~\ref{phi},~\ref{h} and~\ref{a} to upper surface motion equation Eq.~\ref{surface-boundary-condition}, we obtain
\begin{equation}
\begin{split}
\frac{d^2B_{k,n}}{dt^2}+ktanh&(kH)\left(g+\frac{\sigma k^2}{\rho}-Fcos\Omega t\right)B_{k,n}(t)=0\label{parametric-equation}
\end{split}
\end{equation}
It's a parametric resonance equation, describing the response of the drop surface to the vertical excitation $Fcos\Omega t$. Here $k$ can be regarded as a wave-vector of the upper surface. If we denote $p$ and $q$ as follow
\begin{align}
p&=\frac{ktanh(kH)}{\Omega^2}\left(g+\frac{\sigma k^2}{\rho}\right)\\
q&=\frac{kFtanh(kH)}{\Omega^2}
\end{align}
Then Eq.~\ref{parametric-equation} can be transformed to the standard form of Mathieu's equation
\begin{align}
\frac{d^2B_{k,n}}{dt^2}+\left(p+qcos\Omega t\right)B_{k,n}(t)=0\label{Mathieu's-equation}
\end{align}
The natural oscillating frequency $\omega$ of the system can be obtained by setting $\Omega=0$. The principle subharmonic resonance occurs when~\cite{Stability-chart-for-the-delayed-Mathieu-equation}
\begin{align}
\frac{\Omega^2}{4}-\frac{q}{2}\textless\omega^2\textless\frac{\Omega^2}{4}+\frac{q}{2}\label{parametric-condition}
\end{align}
In practical cases, $q$ is much smaller than $p$ so thagt the resonance condition is simplified as $\omega\approx\Omega/2$. When the driving frequency is twice of the system natural frequency, the free surface becomes unstable and the surface oscillation mode increases in amplitude until it is finally restricted by damping or non-linear effects.

On the lateral surface, substituting Eqs.~\ref{phi} and~\ref{a} into motion equation~\ref{periphery-boundary-condition}, we obtain
\begin{align}
\frac{d^2C_{k,n}}{dt^2}+\frac{(n^2-1)k\sigma J'_n(kR)}{\rho R^2J_n(kR)}C_{k,n}(t)=0\label{motion-equation}
\end{align}
Equation~\ref{motion-equation} defines a harmonic oscillating frequency of the drop periphery. 

From~\ref{time-relation-1} and~\ref{time-relation-2} we know that, the induced surface motion due to parametric instability drives azimuthal motion in the same frequency $\omega$ and an azimuthal mode whose eigen frequency is in the vicinity of $\omega$ will be excited. Apparently, the two frequencies defined in Eqs.~\ref{parametric-equation} and~\ref{motion-equation} should be equal, thus we obtain the eigen equation of $k$
\begin{equation}
\begin{split}
\frac{(n^2-1)k\sigma J'_n(kR)}{\rho R^2J_n(kR)}=ktanh(kH)\left(g+\frac{\sigma k^2}{\rho}\right)\label{eigen-equation}
\end{split}
\end{equation}
The eigen value $k$ determines oscillation patterns of the surface. We use $F(k)$ and $G(k)$ to represent the left and right side of the equation~\ref{eigen-equation}, which are plotted together in figure~\ref{Theoretical eigen value calculation}. We find that $F(k)$ increases monotonically with $k$ while $G(k)$ changes quasi-periodically. $F(k)$ and $G(k)$ have multiple intersections, and the $m$ $th$ nonzero intersection determines the $m$ $th$ eigen value $k_m$, which can be solved numerically. $m$ also represents $m$ $th$ eigen mode of the upper surface. The pattern of the water drop is presented in figure~\ref{Surfaces}, in which the height variation $h(r,\theta)$ of the upper surface is denoted with different colors for a $n=6$ drop with $m$ increasing from $1$ to $4$. Along the radial direction, the number of wave nodes is proportional to surface mode number $m$.

%%% figure 3

\begin{figure}[t]
\centering
\includegraphics[width=0.44\textwidth]{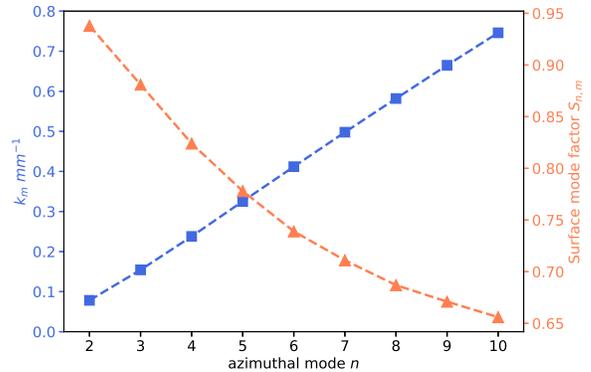}
\caption{Numerical calculation results of both $k_m$ (orange triangles) and surface mode factor $S_{m,n}$ (blue squares). $n$ increases from $2$ to $10$ and we take $m=1$ for all values of $n$. $R=10.8mm$ ($2l_c$), $\sigma=72mN/m$ ($25^{\circ} C$).}
\label{Theoretical eigen values}
\end{figure}

%%% figure 3

From Eqs.~\ref{motion-equation} and~\ref{eigen-equation}, $n$ and $m$ determine the resonance frequency of the water drop, so we have the dispersion relation
\begin{align}
\omega_{m,n}&=\sqrt{\frac{(n^2-1)k_m\sigma J'_n(k_m R)}{\rho R^2 J_n(k_m R)}}\label{new-dispersion-relation}
\end{align}
When the surface motion is ignored, i.e. $k=0$, we have the following equation
\begin{align}
\lim_{k \to 0}\frac{kJ'_n(kR)}{J_n(kR)}=\frac{n}{R}\label{limit-relation}
\end{align}
Then equation~\ref{new-dispersion-relation} naturally reduces to Rayleigh equation~\ref{Rayleigh-equation}. We define a factor $S_{n,m}$
\begin{align}
S_{n,m}=\frac{\omega_{n,m}^2}{\omega_{n}^2}=\frac{R}{n}\frac{k_m J'_n(k_m R)}{J_n(k_m R)}
\end{align}
It reflects the influence of the surface mode on the resonance frequency.

%%% figure 4

\begin{figure}[t]
\centering

\includegraphics[width=0.44\textwidth]{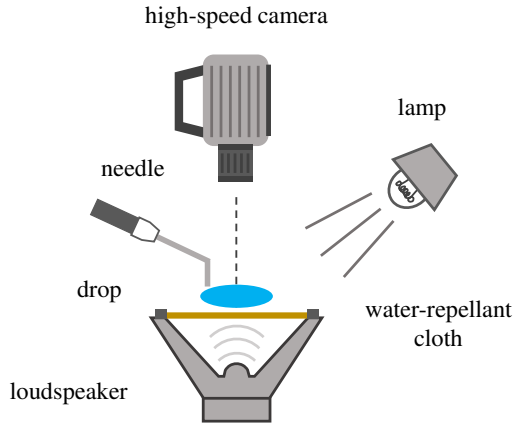}

\caption{Sketch of the experimental setup. The loudspeaker is driven by a signal generator. The needle places water drops on the hydrophobic cloth. The loudspeaker outputs low-frequency sound waves to excite the drops into oscillations. A high speed camera records the top view images.}
\label{Facilities}
\end{figure}

%%% figure 4

In figure~\ref{Theoretical eigen value calculation}, we can find that $m=1$ and higher modes $m$ have different oscillation frequencies. In practical cases, each oscillation frequency $\omega$ should correspond to one mode $m$. Usually, for the upper surface of the drop, higher-order modes are suppressed due to the dissipation of the system and an appreciable disturbance at the surface occurs only when a low-order mode is excited~\cite{The-Royal-Society-The-stability-of-the-plane-free-surface-of-a-liquid-in-vertical-periodic-motion}. We numerically calculate $k_m$ and $S_{m,n}$ for $m=1$ and with $n$ increasing from $2$ to $10$. As shown in figure~\ref{Theoretical eigen values}, for drops with the same radius $R=10.8mm$, the larger $n$, the larger the wave vector $k_m$, and the smaller the factor $S_{n,m}$ (all of them are smaller than $1$), which corresponds to a more significant softening to resonance frequency.

For an oscillation system, the oscillating frequency is determined by the stiffness coefficient. The existence of surface modes would decrease the surface tension potential, consequently decreases the stiffness coefficient for the oscillation. Ignoring surface modes and reducing the motion of the water drop to two-dimensional overestimates the potential energy as well as the resonance frequency. We will compare the frequency to experimental results in the following section. And the correction also explains the discrepancy of oscillation frequency found in Ref.~\cite{PRE-Oscillating-and-star-shaped-drops-levitated-by-an-airflow}.

%%% figure 5

\begin{figure}[t]
\centering

\subfigure[n=3]{
\begin{minipage}[t]{0.14\textwidth}
\centering
\includegraphics[width=1\textwidth,height=1\textwidth]{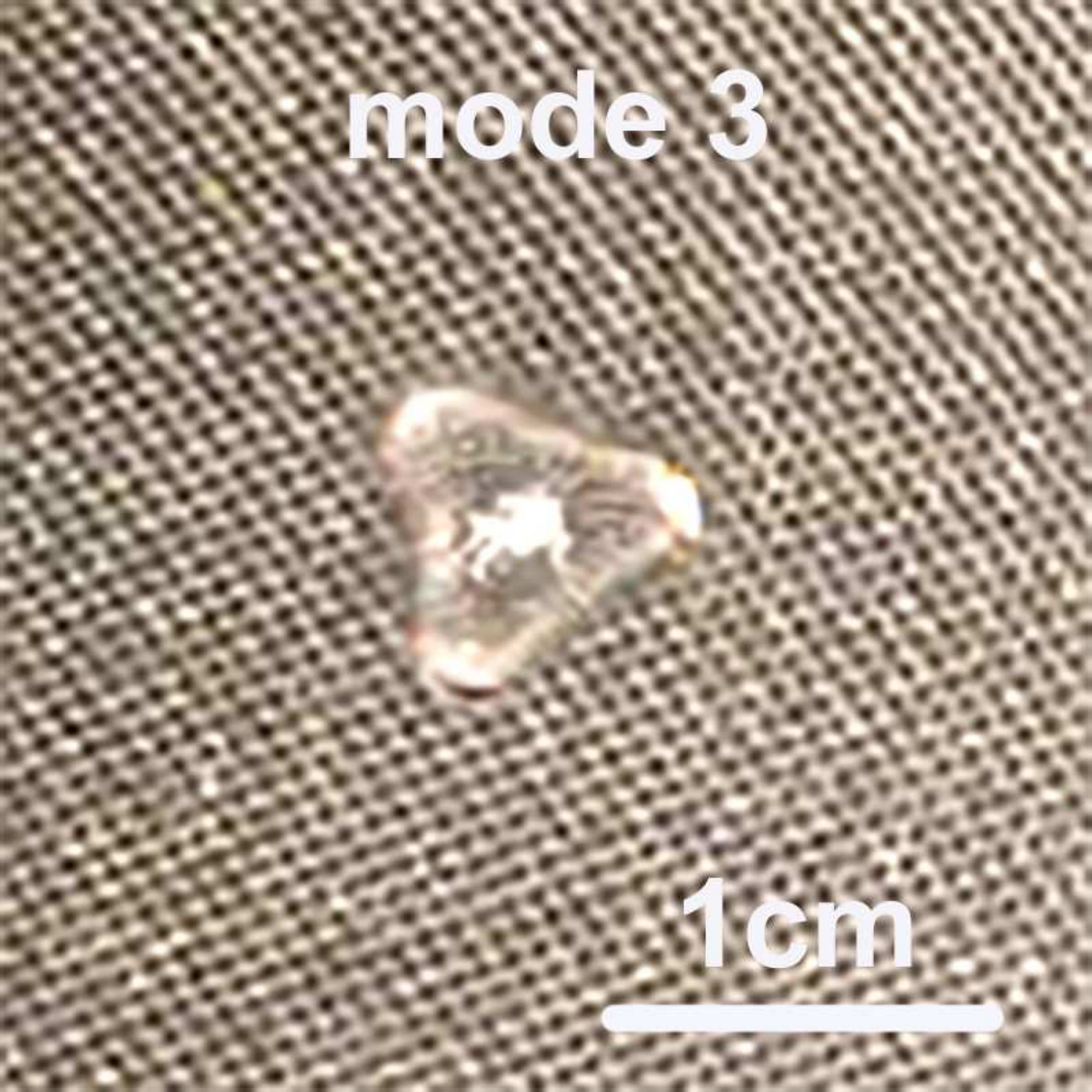}
\end{minipage}
}
\subfigure[n=4]{
\begin{minipage}[t]{0.14\textwidth}
\centering
\includegraphics[width=1\textwidth,height=1\textwidth]{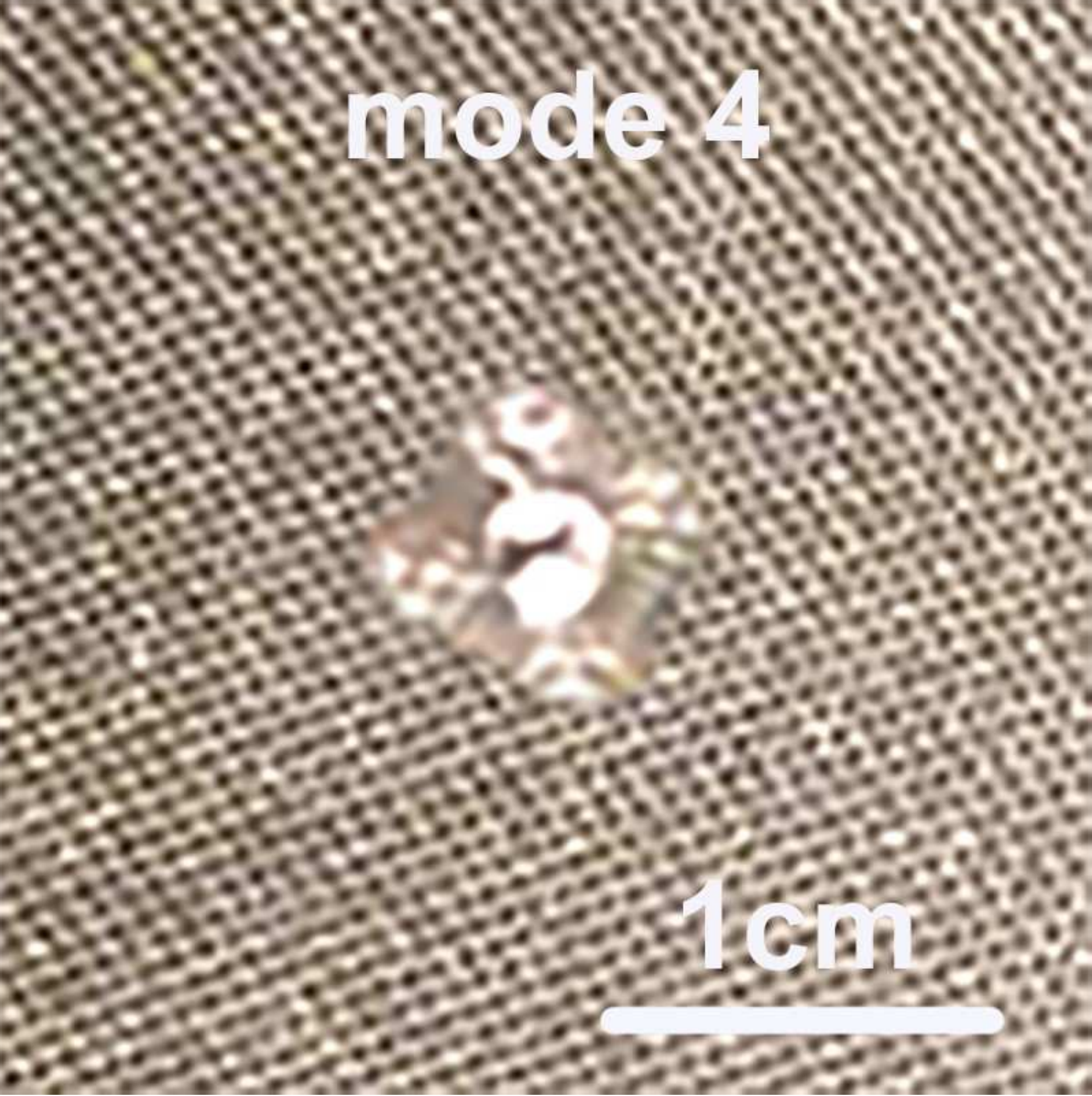}
\end{minipage}
}
\subfigure[n=5]{
\begin{minipage}[t]{0.14\textwidth}
\centering
\includegraphics[width=1\textwidth,height=1\textwidth]{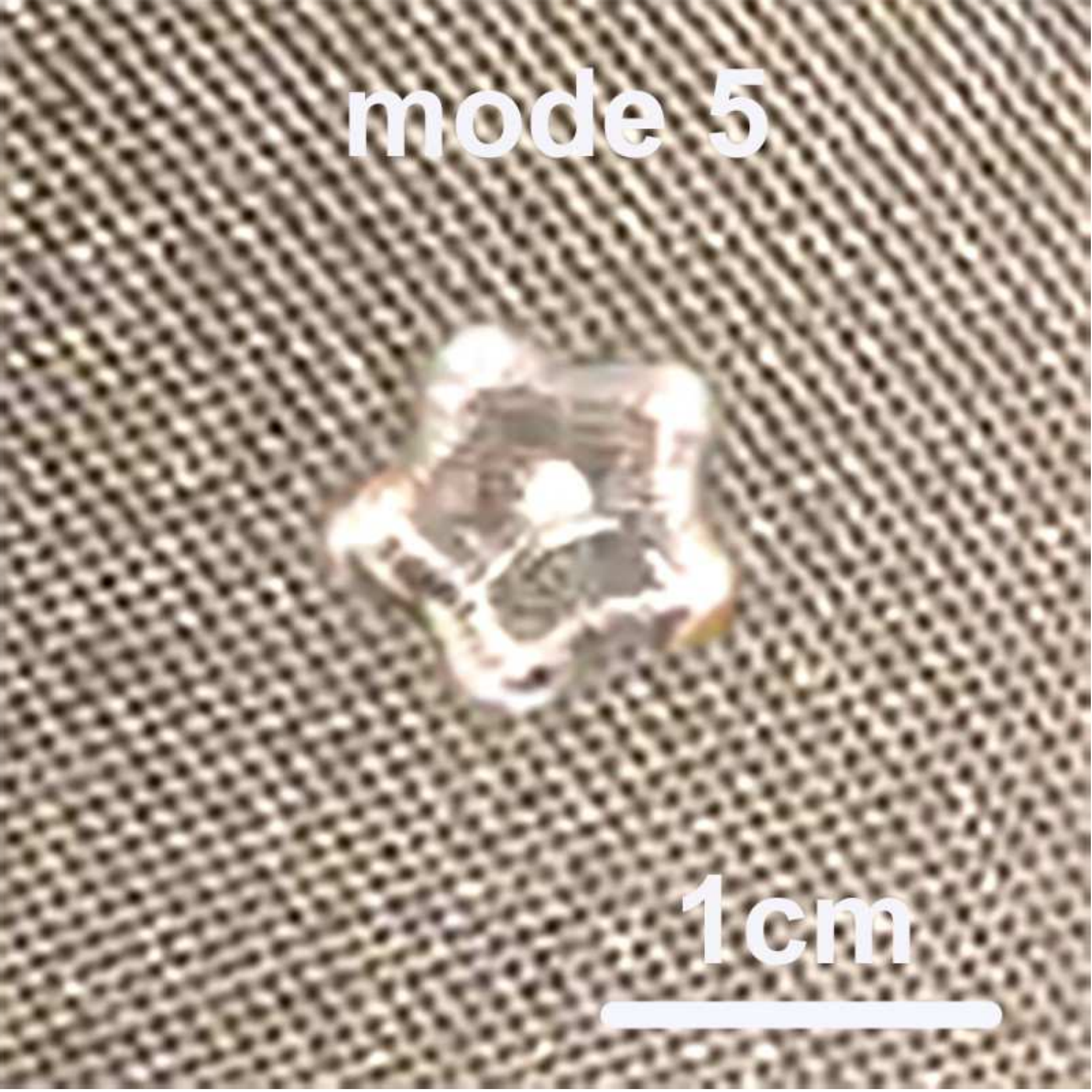}
\end{minipage}
}

\subfigure[n=6]{
\begin{minipage}[t]{0.14\textwidth}
\centering
\includegraphics[width=1\textwidth,height=1\textwidth]{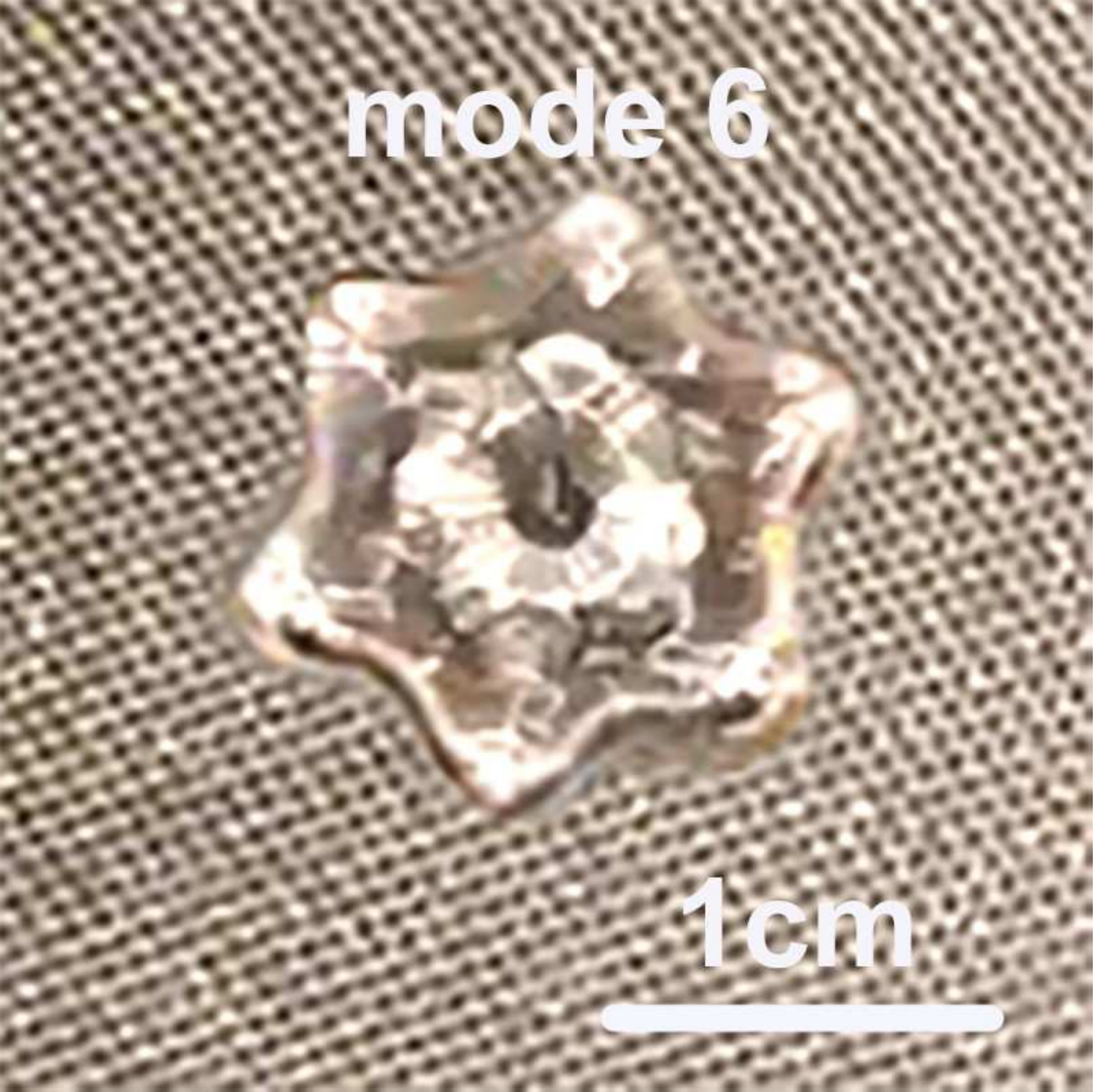}
\end{minipage}
}
\subfigure[n=7]{
\begin{minipage}[t]{0.14\textwidth}
\centering
\includegraphics[width=1\textwidth,height=1\textwidth]{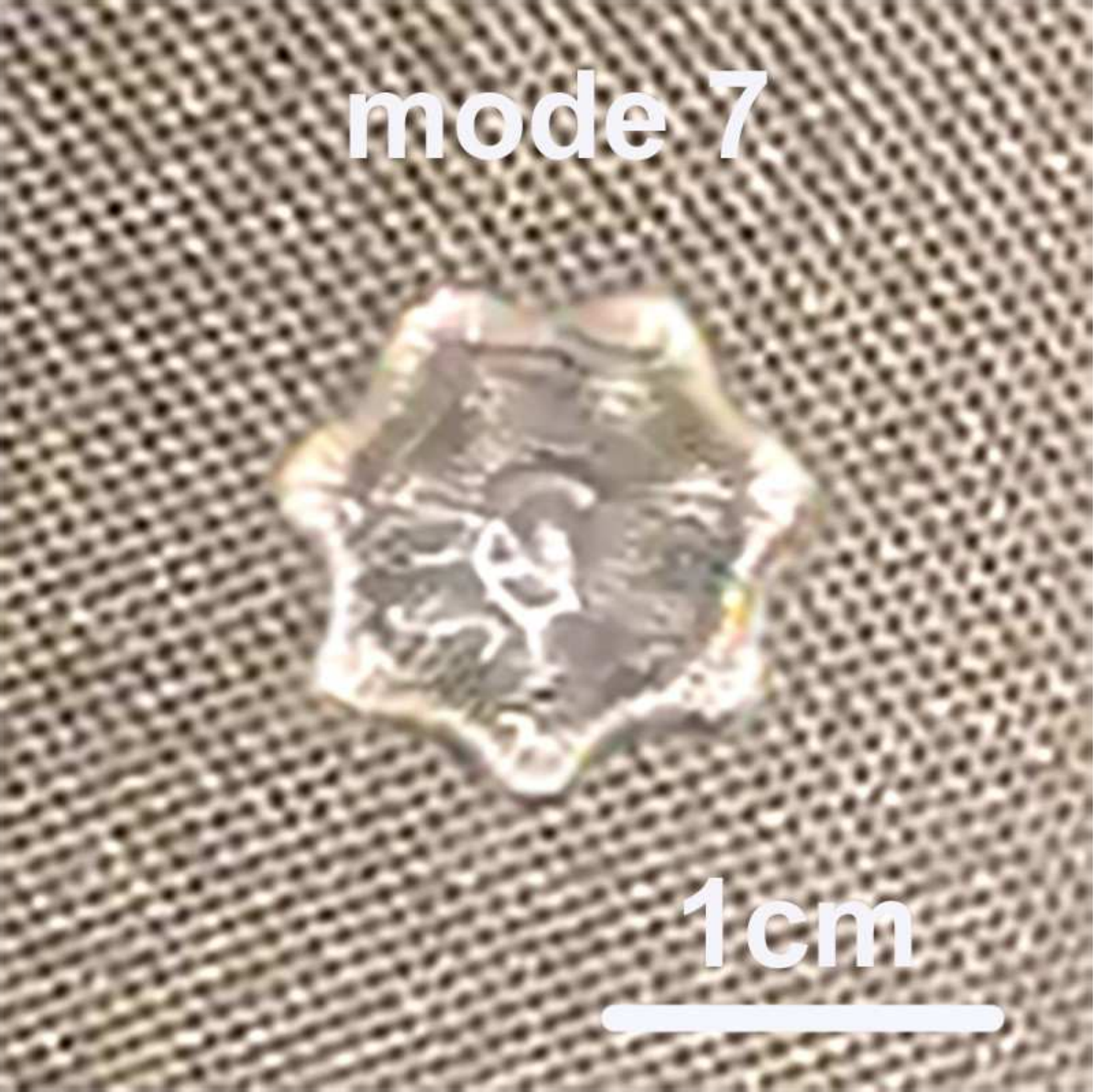}
\end{minipage}
}
\subfigure[n=8]{
\begin{minipage}[t]{0.14\textwidth}
\centering
\includegraphics[width=1\textwidth,height=1\textwidth]{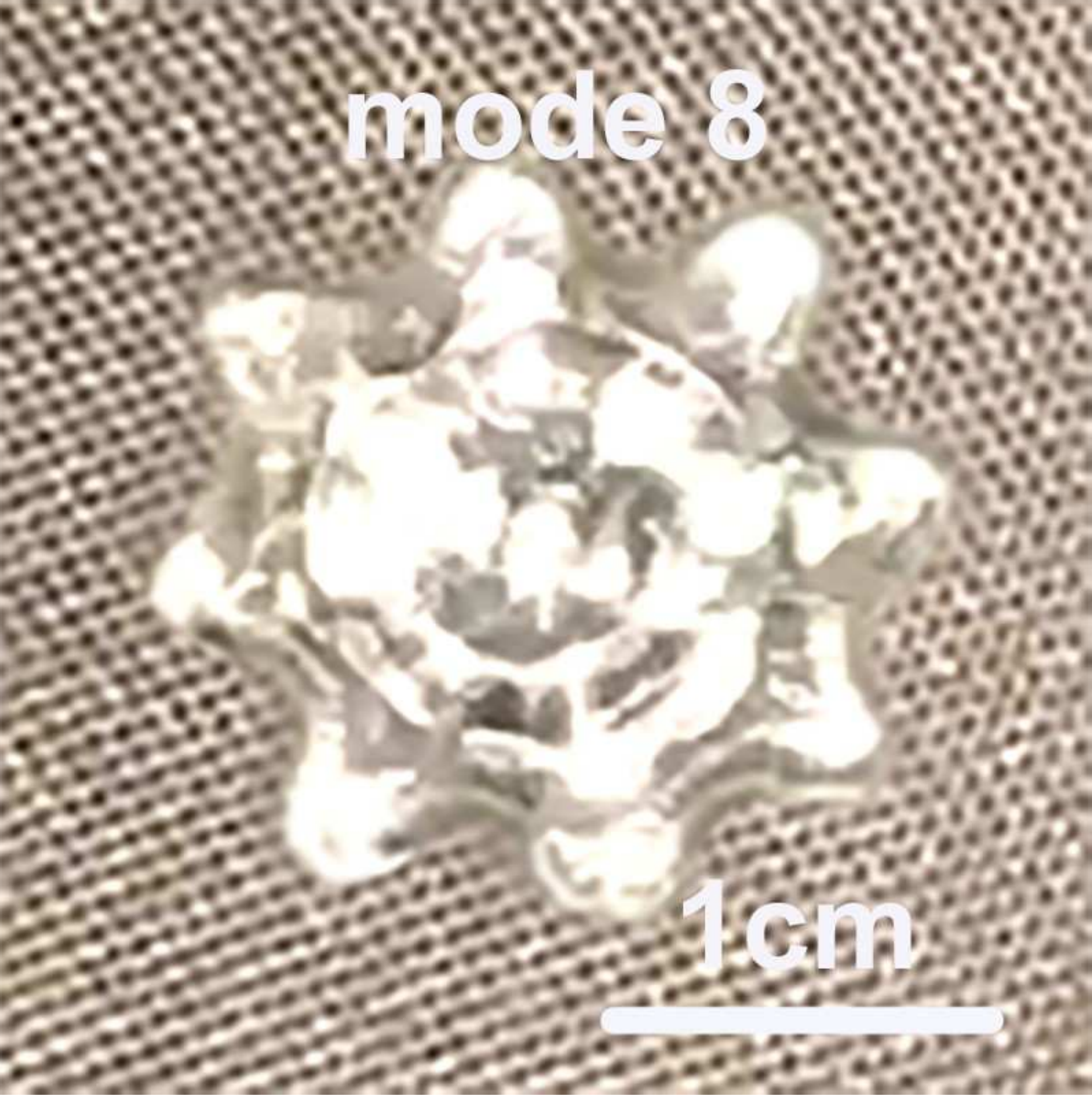}
\end{minipage}
}

\subfigure[n=9]{
\begin{minipage}[t]{0.14\textwidth}
\centering
\includegraphics[width=1\textwidth,height=1\textwidth]{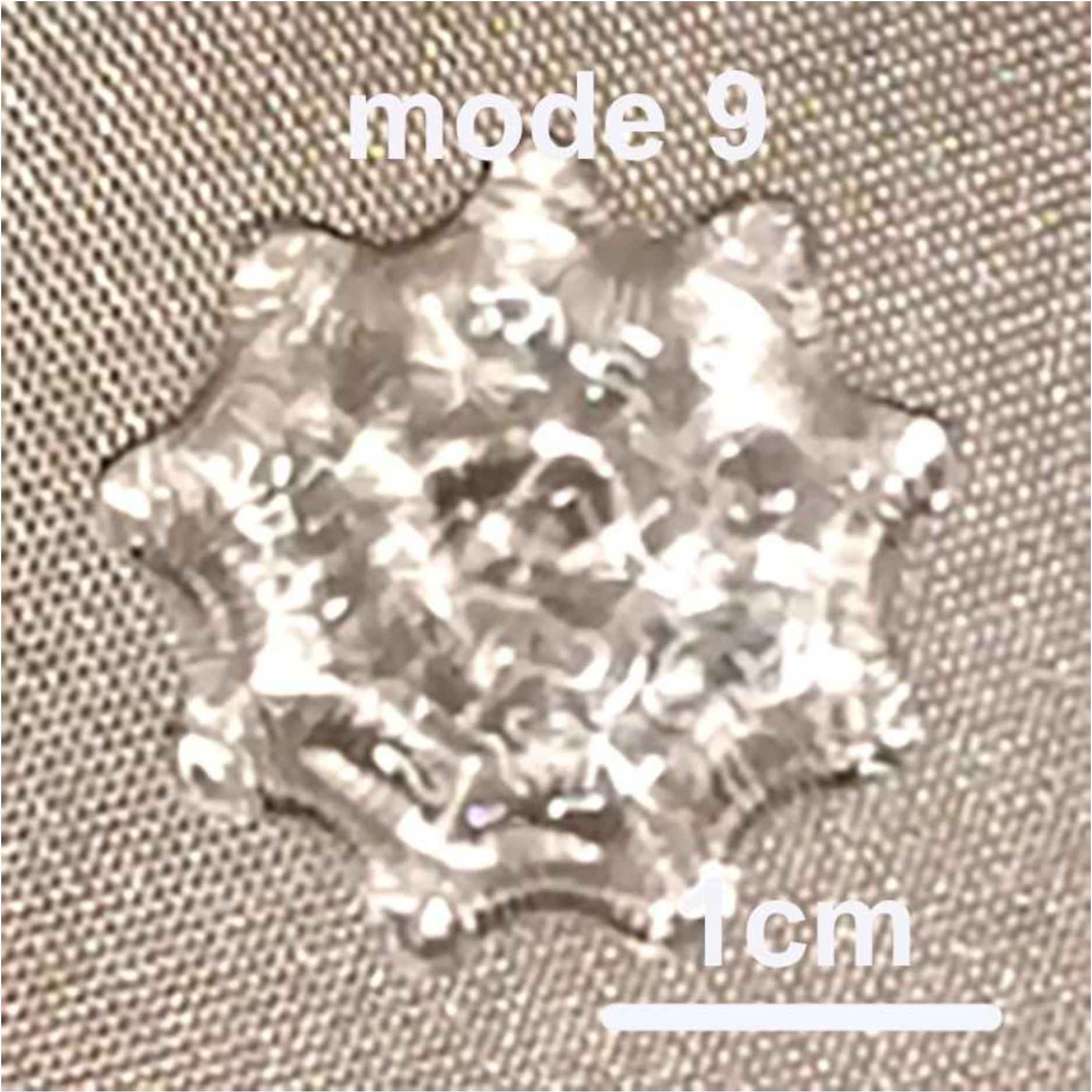}
\end{minipage}
}
\subfigure[n=10]{
\begin{minipage}[t]{0.14\textwidth}
\centering
\includegraphics[width=1\textwidth,height=1\textwidth]{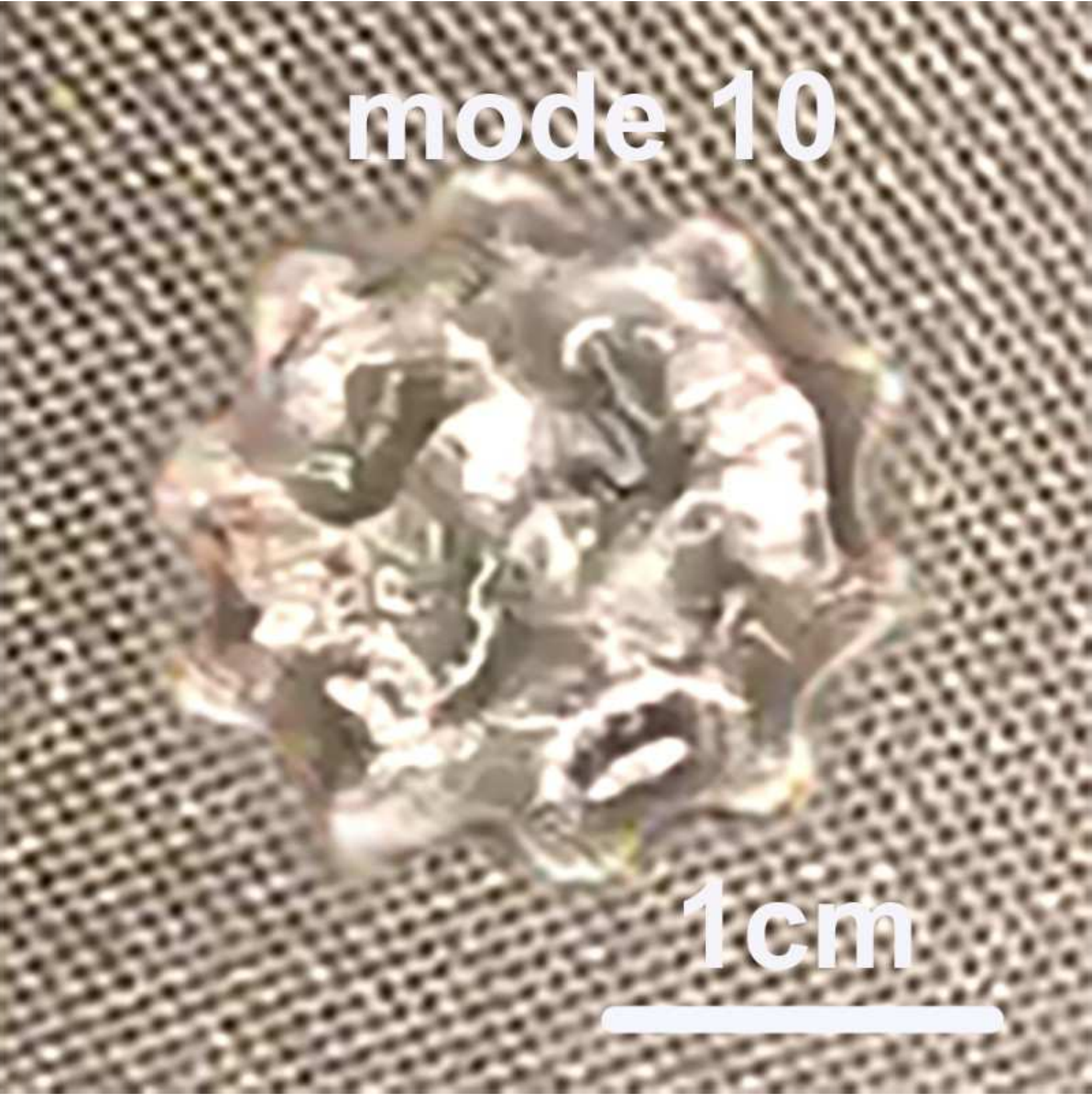}
\end{minipage}
}
\subfigure[n=11]{
\begin{minipage}[t]{0.14\textwidth}
\centering
\includegraphics[width=1\textwidth,height=1\textwidth]{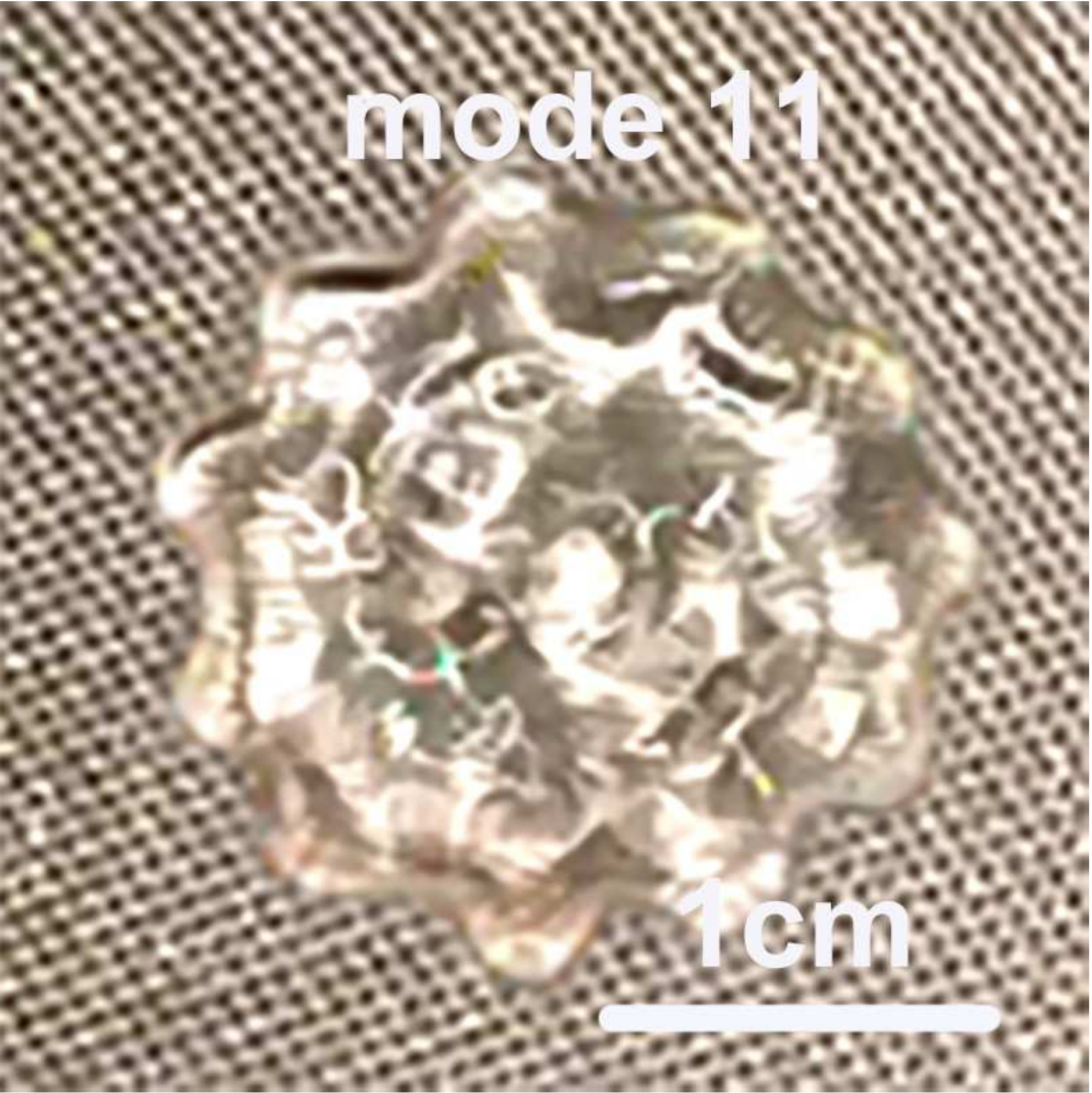}
\end{minipage}
}

\caption{Examples of instantaneous top views of star-shaped oscillating drops. From (a) to (i) the values of $n$ increase from $3$ to $11$.}
\label{Experimental observations}
\end{figure}

%%% figure 5

%%% figure 6

\begin{figure}[t]
\centering

\subfigure[Experimental]{
\begin{minipage}[t]{0.18\textwidth}
\centering
\includegraphics[width=1\textwidth,height=1\textwidth]{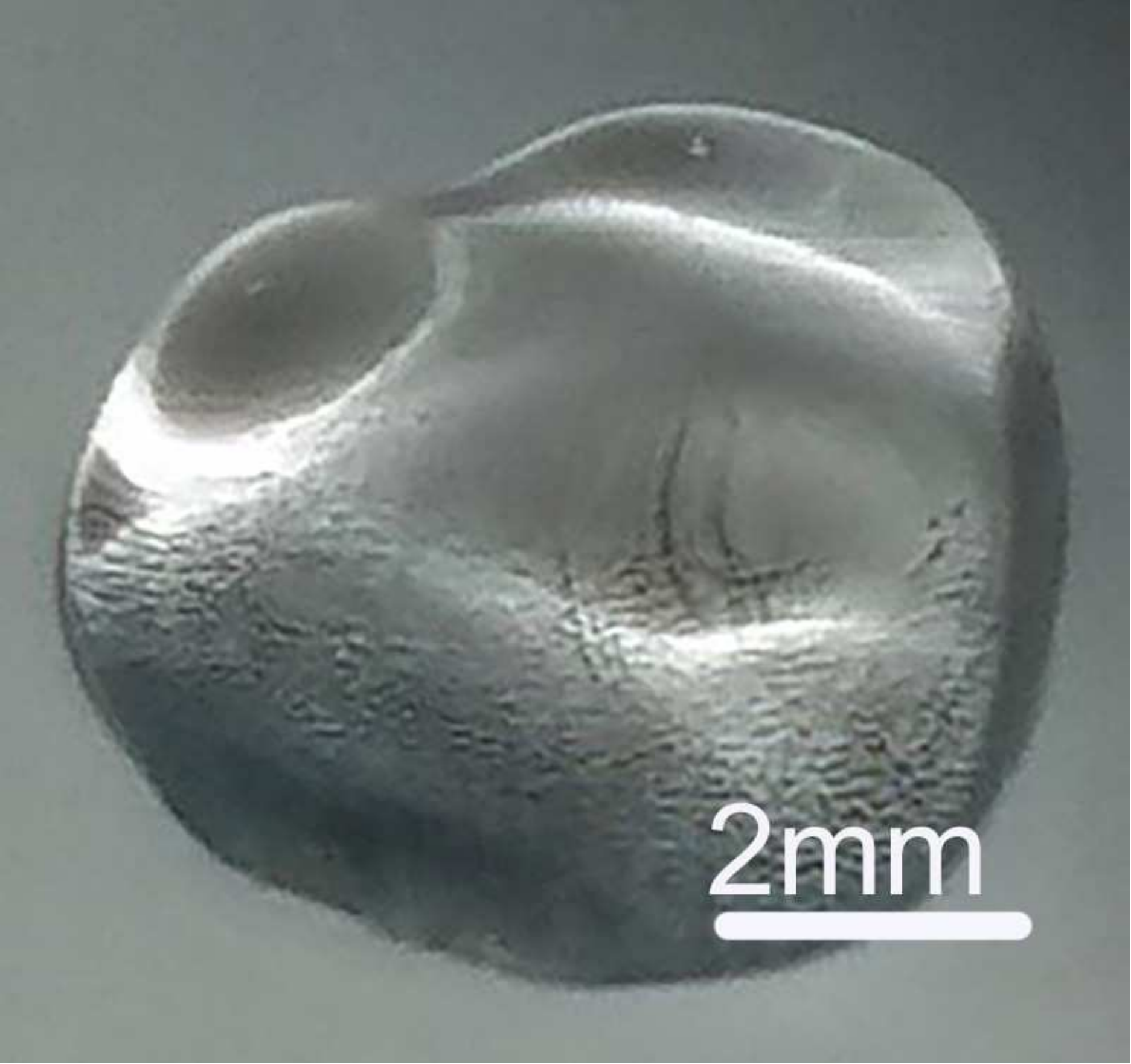}
\end{minipage}
}
\subfigure[Theoretical]{
\begin{minipage}[t]{0.18\textwidth}
\centering
\includegraphics[width=1.1\textwidth,height=1\textwidth]{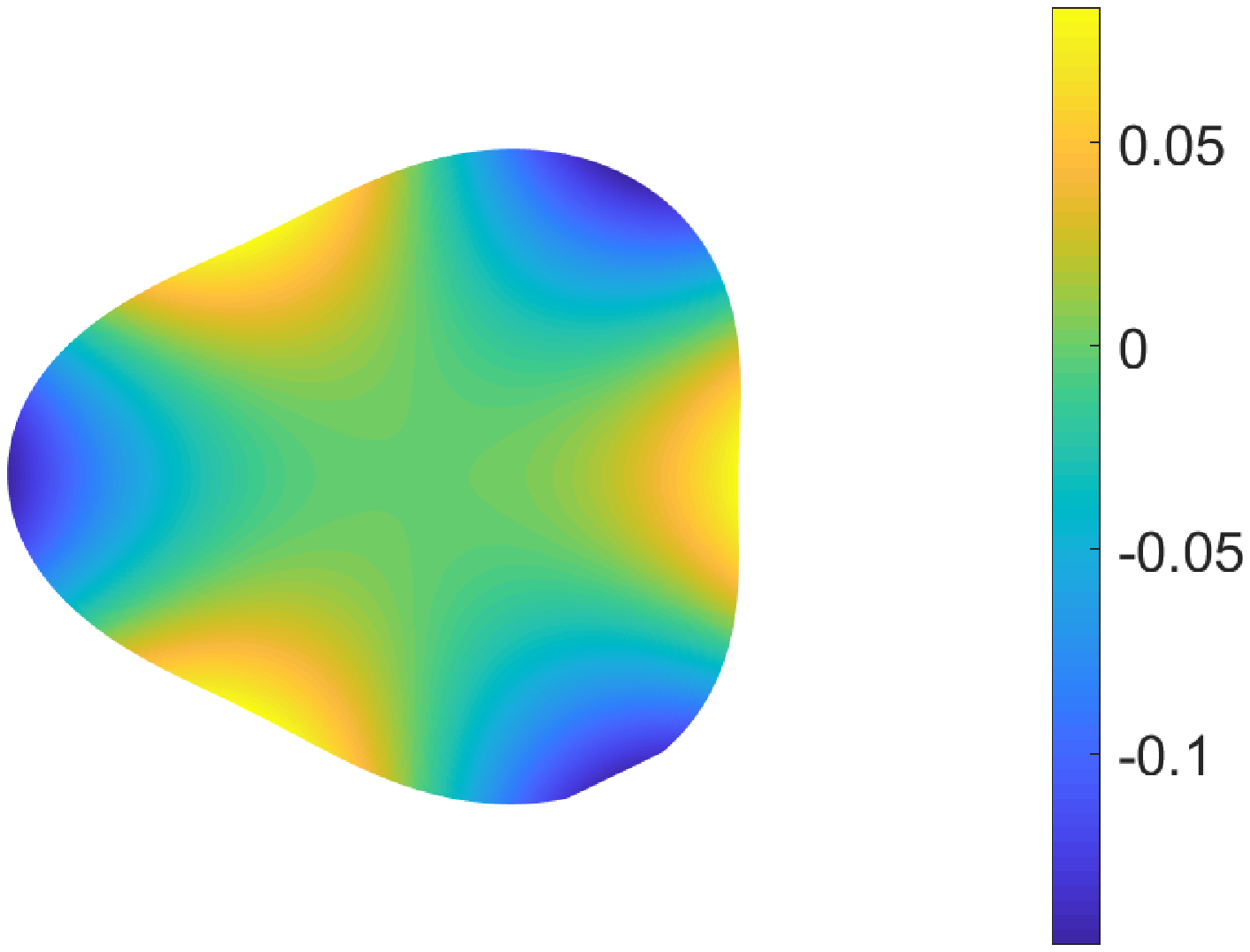}
\end{minipage}
}

\caption{(a) The surface pattern of a $n=3$ drop captured from very close distance. We have enhanced the image contrast for visibility. (b) Corresponding theoretical surface mode ($m=1$). The numerically solved wave-vector $k_1$ is $0.99mm^-1$. The height variation is denoted with different colors. The colorbar
represents height (in $mm$).}
\label{Surface mode comparison}
\end{figure}

%%% figure 6

\section{Experiment and results}\label{Experiment}

%%% figure 7

\begin{figure*}[t]
\centering

\subfigure[60Hz]{
\begin{minipage}[t]{0.31\textwidth}
\centering
\includegraphics[width=1\textwidth,height=0.6\textwidth]{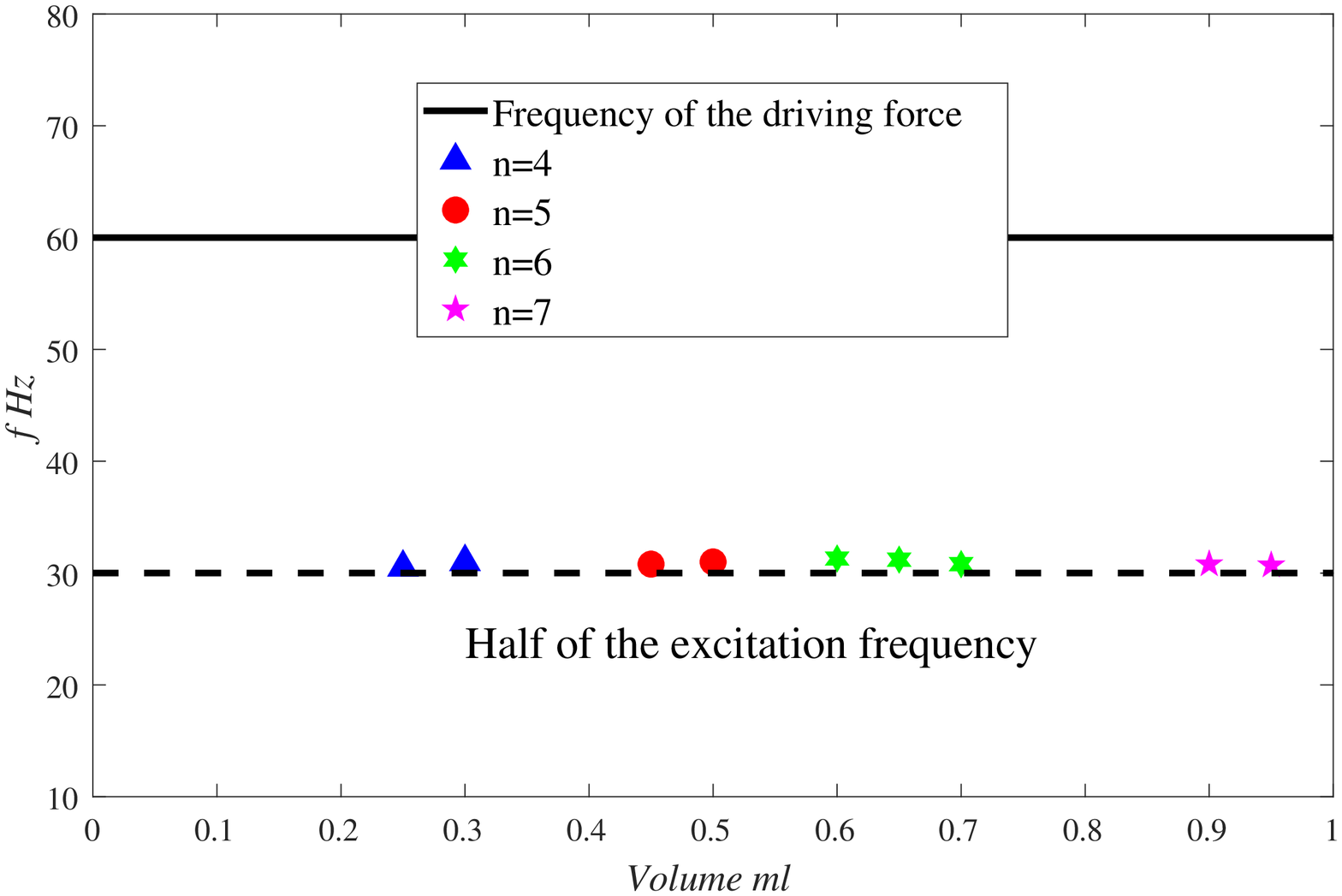}
\end{minipage}
}
\subfigure[90Hz]{
\begin{minipage}[t]{0.31\textwidth}
\centering
\includegraphics[width=1\textwidth,height=0.6\textwidth]{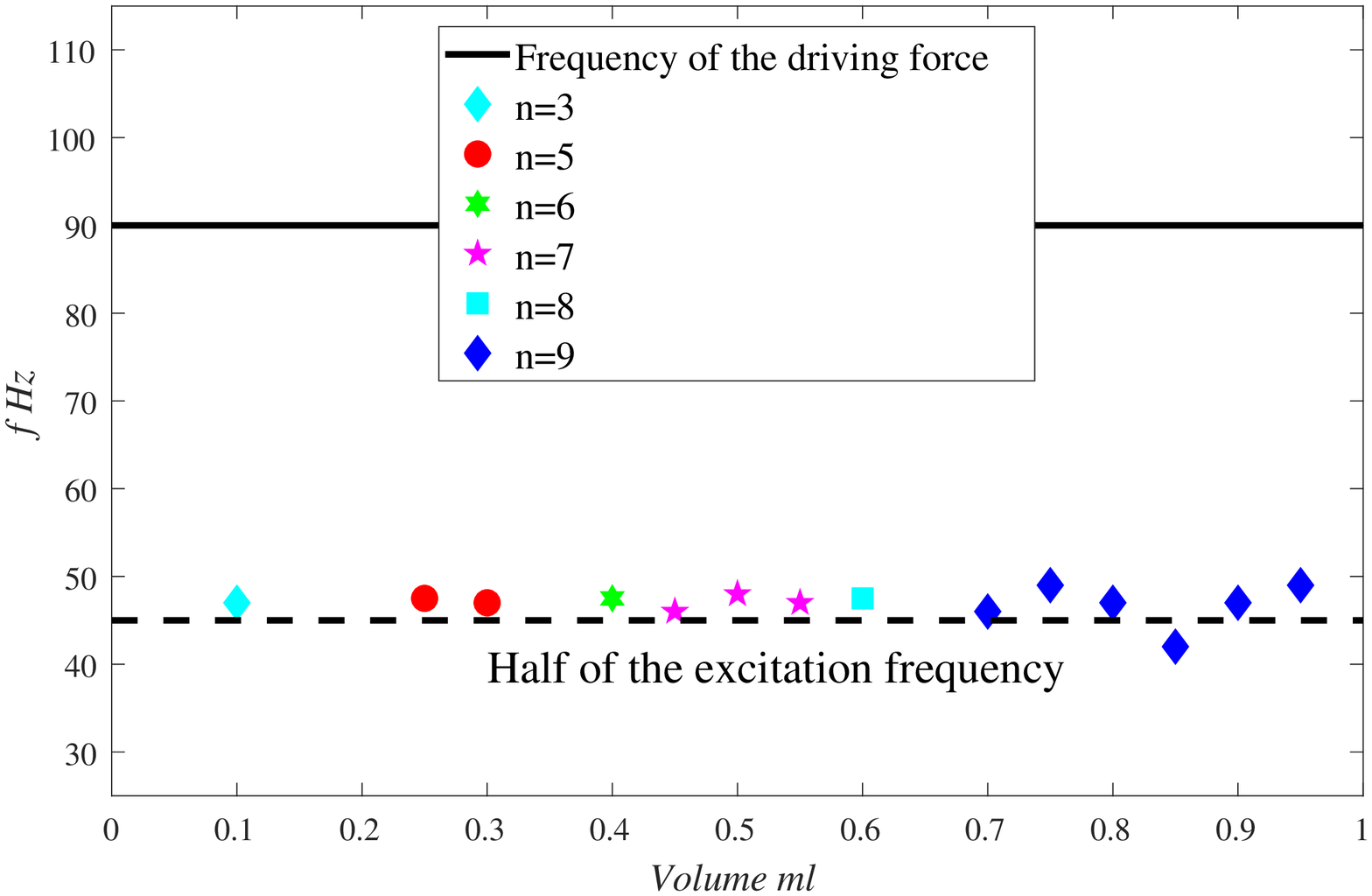}
\end{minipage}
}
\subfigure[120Hz]{
\begin{minipage}[t]{0.31\textwidth}
\centering
\includegraphics[width=1\textwidth,height=0.6\textwidth]{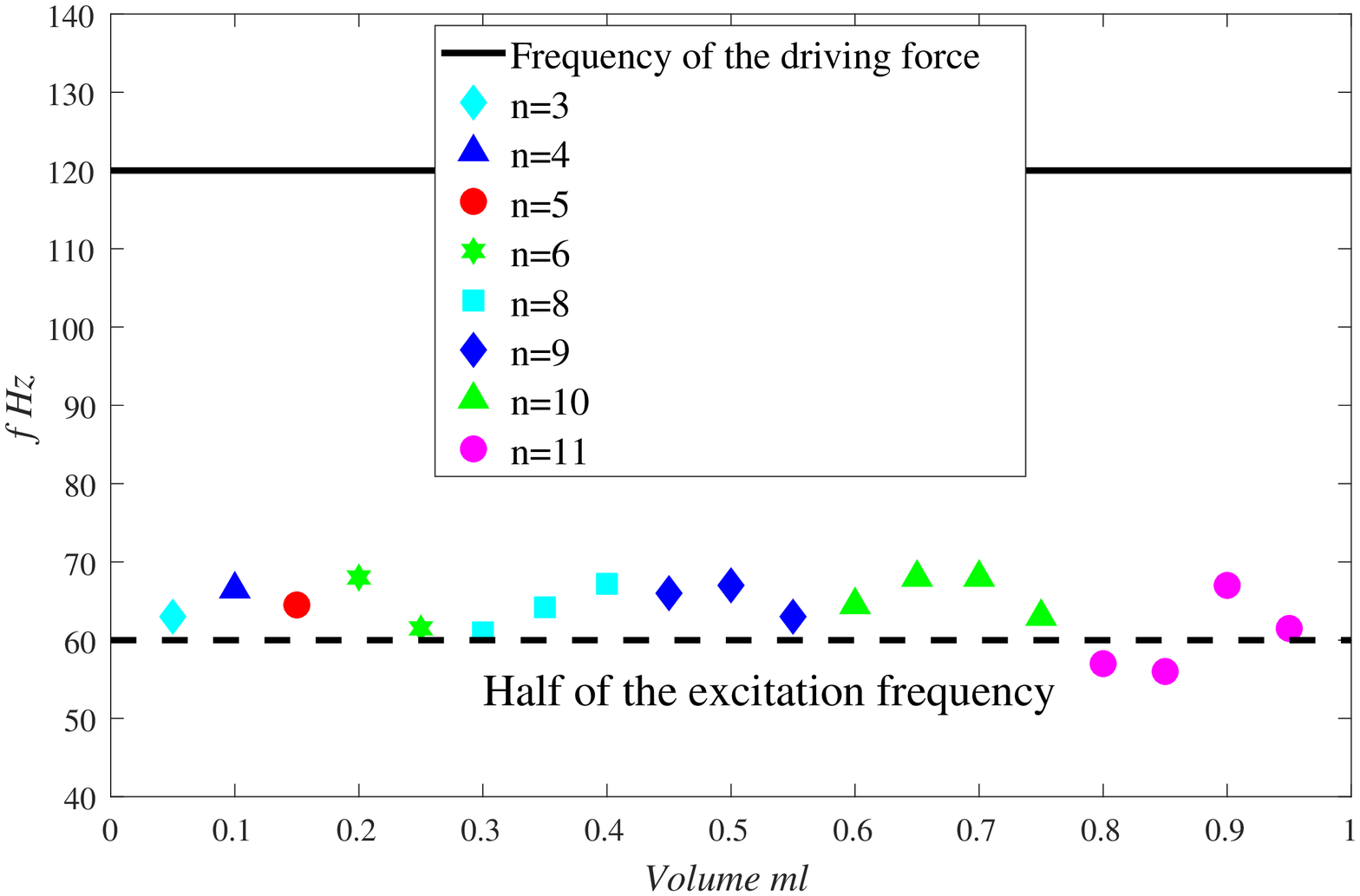}
\end{minipage}
}

\caption{Relation between the measured oscillation frequency $f$ of star-shaped drops and half of the excitation frequency $f_0/2$. From (a) to (c) $f_0$ are $60Hz$, $90Hz$, and $120Hz$ respectively.}
\label{Relation between measure and half of the excitation}
\end{figure*}

%%% figure 7

%%% figure 8

\begin{figure*}[t]
\centering

\subfigure[60Hz]{
\begin{minipage}[t]{0.31\textwidth}
\centering
\includegraphics[width=1\textwidth,height=0.7\textwidth]{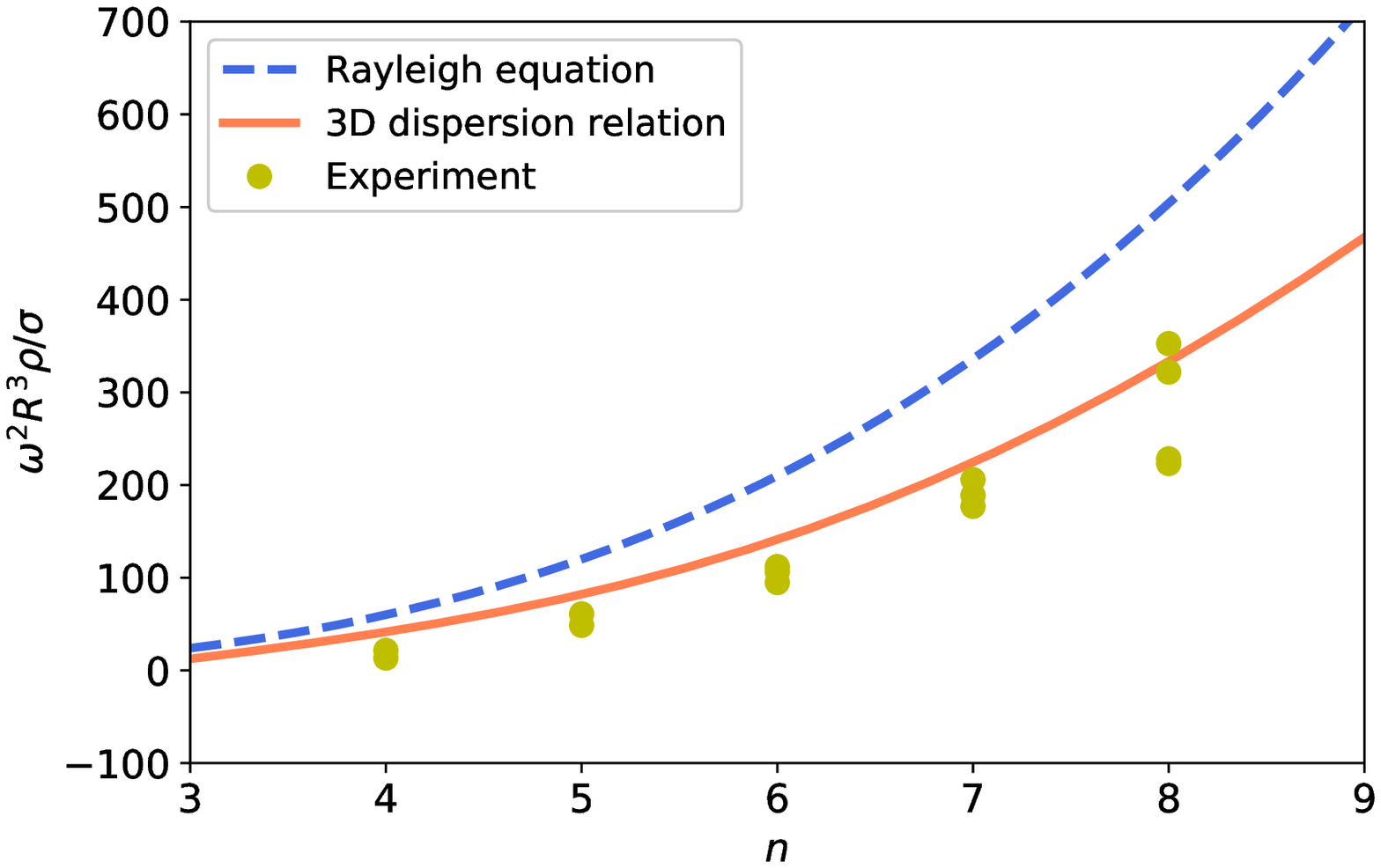}
\end{minipage}
}
\subfigure[90Hz]{
\begin{minipage}[t]{0.31\textwidth}
\centering
\includegraphics[width=1\textwidth,height=0.7\textwidth]{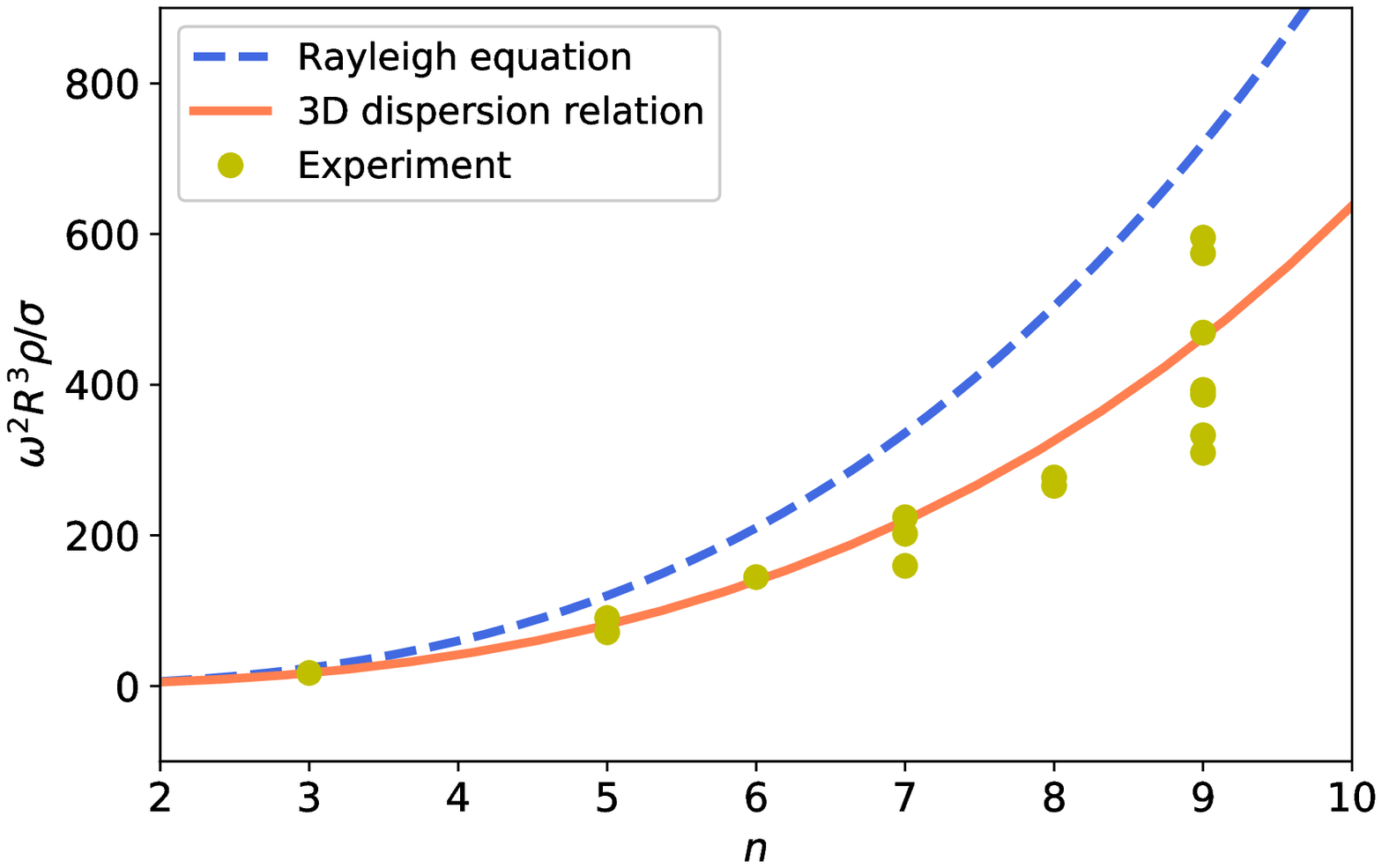}
\end{minipage}
}
\subfigure[120Hz]{
\begin{minipage}[t]{0.31\textwidth}
\centering
\includegraphics[width=1\textwidth,height=0.7\textwidth]{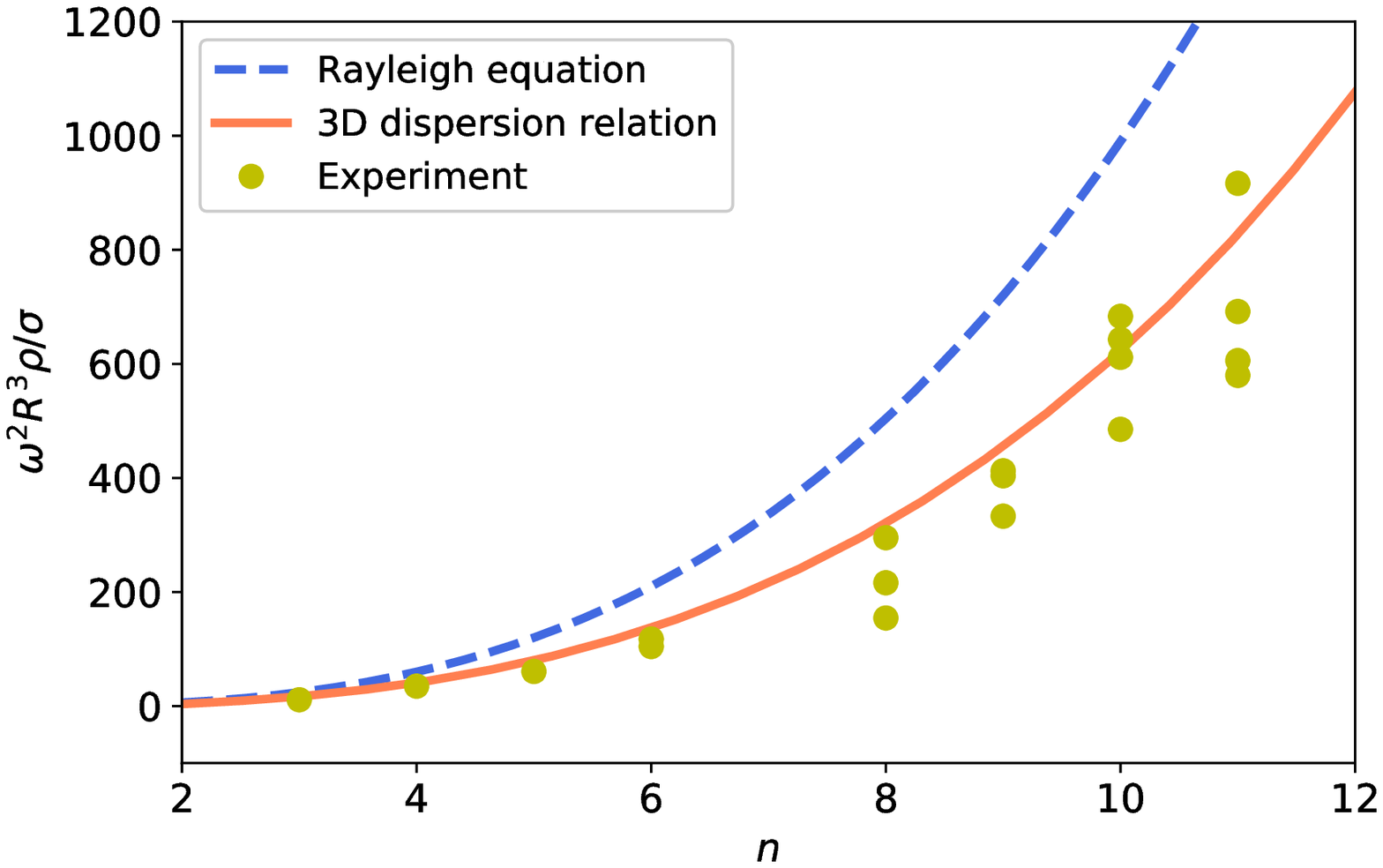}
\end{minipage}
}

\caption{Theoretically predicted values of $\omega^2 R^3\sigma /\rho$ and experimental values. Blue dashed lines represent the values obtained from Rayleigh equation. Orange solid lines represent those obtained from the 3D dispersion relation (the data is calculated for each azimuthal mode $n$ the line is connected using spline interpolation). Yellow dots are experimental values. From (a) to (c) $f_0$ are $60Hz$, $90Hz$, and $120Hz$ respectively.}
\label{Experimental comparison}
\end{figure*}

%%% figure 8

Experimental setup is shown in Figure~\ref{Facilities}. A water-repellent cloth is attached horizontally to a loudspeaker cone (So-Voioe SVF149WR). For each drop the measured contact angle is more than $120^\circ$ so that the hysteretic behavior can be avoided. The loudspeaker is connected to a signal generator (Right SG1020P), which applies a vertical excitation to the drop. We use a injector to control the volume and place the water drop on the cloth. The top view of the water drop is recorded using a high-speed video camera (Metalab 300C-U3) at the rate of 400 frames/s. When a drop performs stationary star-shaped oscillation, the camera records a sequence of images. The oscillating frequency $f$ and mean radius $R$ of the drop are analyzed frame by frame.

We set the input sinusoidal waves at the driving frequency $f_0=\Omega/2\pi=60Hz,90Hz,120Hz$ respectively. The volume of the water drop is increased by $0.05ml$ each time, measured by the injector. For $f_0=120Hz$, star-shaped oscillating drops from $n=3$ to $n=11$ are observed, as shown in figure~\ref{Experimental observations}. We can see the petal-like patterns at the upper surface, illustrating the existence of surface motion patterns. For $f_0=90Hz$, star-shaped drops $n=3,5,6,7,8,9$ are observed; for $f_0=60Hz$, star-shaped drops $n=4,5,6,7$ are observed. The surface mode computed with our model is compared to the experimental photograph for a $n=3$ drop in figure~\ref{Surface mode comparison}. The radius of the drop is $2.7mm$ and the calculated wave-vector $k_1$ is $0.99mm^-1$. The drop is captured from very close distance and we enhance the image contrast for visibility. The theoretical image resembles the experimental one. We show the relation between the measured oscillating frequency $f$ and half of the excitation frequency $f_0/2$ in figure~\ref{Relation between measure and half of the excitation}. The subharmonic parametric resonance condition~\ref{parametric-condition} is satisfied. 
%Additionally, we find that for larger $f_0$ the resonance oscillation frequency shows a slight deviation from $f_0/2$, caused by the increase of parameter $q$ as well as the subharmonic resonance domain.

In dispersion relation~\ref{Rayleigh-equation}, the predicted value of $\omega^2 $ varies as $n(n^2 -1)$, while in the new dispersion relation~\ref{new-dispersion-relation} we propose, it varies as $n(n^2 -1)$ multiplied by an additional factor $S_{n,m}$. For each azimuthal mode $n$, we calculate the eigen value of $k_m$ by solving equation~\ref{eigen-equation} numerically and we take mode number $m=1$. For $f_0=120Hz$,$90Hz$,$60Hz$, $k_1$ are calculated and equal to $1.4mm^{-1}$,$1.0mm^{-1}$,$0.8mm^{-1}$ respectively. We find $k_1$ appears to be independent of the azimuthal mode $n$, only determined by the excitation frequency $f_0$. In figure~\ref{Experimental comparison} we plot the theoretical results according to~\ref{Rayleigh-equation} and~\ref{new-dispersion-relation}, as well as the experimental results of $\omega^2 R^3\sigma /\rho$. The yellow dots are experimental values. The blue dashed lines represent the values predicted by~\ref{Rayleigh-equation}, obviously overestimating the oscillating frequency. And the orange solid lines represent those predicted by~\ref{new-dispersion-relation}, which fits the experimental data much better. For some azimuthal number $n$ the observed oscillation frequency consist of a set of data points. Actually these drops have the same azimuthal mode $n$ but are slightly different in radius $R$ and surface factor $S_{n,m}$. while we only substitute in~\ref{new-dispersion-relation} the average R observed in experiment.

\section{Conclusion}\label{Conclusion}

In this paper, we derive a complete theoretical model to include both surface modes and azimuthal modes of water drops under vertical excitation. The model we propose is applicable to drops under any form of vertical excitation. We prove that the star-shaped oscillation originates from parametric instability of the upper surface and explain the mechanism that leads to the star-shaped oscillation. We also propose a new dispersion relation based on the combination of surface modes and azimuthal modes, which gives a lower resonance frequency due to the additional surface mode. The dispersion relation explains the discrepancy of oscillation frequencies found in previous studies and is in good agreement with our experimental data. These results enhance our understanding of the dynamics of water drops.

\section*{Acknowledgements}

The authors are grateful to Mr. Y. Luo for beneficial discussions.

%\bibliographystyle{plain}
%\bibliographystyle{unsrt}
%\bibliography{refer}

\end{document}